\documentclass[aps,prd,nofootinbib,onecolumn,superscriptaddress,preprintnumbers,balancelastpage,longbibliography,nobibnotes]{revtex4-2}

\usepackage{hyperref}
\usepackage{amsfonts}
\usepackage{mathrsfs}
\usepackage{epsfig}
\usepackage{graphicx}            
\usepackage{url}
\usepackage{hyperref}
\usepackage{float}
\usepackage{subcaption, caption}\usepackage{xargs} 
\usepackage{color}
\usepackage{todonotes}
\usepackage{amsmath, amssymb}

\newcommandx{\task}[2][1=]{\todo[size=\small,linecolor=red,backgroundcolor=red!25, bordercolor=red,#1]{#2}}
\newcommandx{\change}[2][1=]{\todo[size=\small,linecolor=blue,backgroundcolor=blue!25,bordercolor=blue,#1]{#2}}
\usepackage{graphicx}
\usepackage{epsf}

\newcommand{\be}{\begin{equation}}
\newcommand{\ee}{\end{equation}}
\newcommand{\bea}{\begin{eqnarray}}
\newcommand{\eea}{\end{eqnarray}}

\usepackage{comment}

\newcommand{\nc}{\newcommand}

\nc{\beq}{\begin{equation}}
\nc{\eeq}{\end{equation}}
\nc{\beqa}{\begin{eqnarray}}
\nc{\eeqa}{\end{eqnarray}}

\usepackage{slashed}

\newcommand{\lsim}{\!\mathrel{\hbox{\rlap{\lower.55ex \hbox{$\sim$}} \kern-.34em \raise.4ex \hbox{$<$}}}}
\newcommand{\gsim}{\!\mathrel{\hbox{\rlap{\lower.55ex \hbox{$\sim$}} \kern-.34em \raise.4ex \hbox{$>$}}}}

\def\be{\begin{equation}}
\def\ee{\end{equation}}

\newcommand{\Fref}[1]{Fig.\,\ref{#1}}
\newcommand{\Eref}[1]{Eq.\,(\ref{#1})}

\newcommand\affspc{\vspace{4pt}}

\usepackage{footmisc}
\usepackage{setspace}

\begin{document}

\preprint{DESY-23-121, TTP23-033, P3H-23-056}

\title{On Particle Production from Phase Transition Bubbles}

\author{Henda Mansour}
\affiliation{Deutsches Elektronen-Synchrotron DESY, Notkestr.\,85, 22607 Hamburg, Germany \affspc}
\affiliation{II. Institute of Theoretical Physics, Universität Hamburg, 22761, Hamburg, Germany \affspc}
\affiliation{Institute for Theoretical Particle Physics, Karlsruhe Institute of Technology, 76131, Karlsruhe, Germany \affspc}
\author{Bibhushan Shakya}
\affiliation{Deutsches Elektronen-Synchrotron DESY, Notkestr.\,85, 22607 Hamburg, Germany \affspc}

\begin{abstract}
While first order phase transitions (FOPTs) have been extensively studied as promising cosmological sources of gravitational waves, the phenomenon of particle production from the dynamics of the background field during FOPTs has received relatively little attention in the literature, where it has only been studied with semi-analytic estimates in some simplified settings. This paper provides improved numerical studies of this effect in more realistic frameworks,  revealing important qualitative details that have been missed in the literature. We also provide easy to use analytic formulae that can be used to calculate particle production in generic FOPT setups.

\end{abstract}

\maketitle

\tableofcontents

\newpage

\section{Introduction}

The physics of a first order phase transition (FOPT), where the false (metastable) vacuum of a theory decays into the energetically favored true (stable) vacuum through the nucleation and collision of true vacuum bubbles \cite{Hogan:1983ixn,Witten:1984rs,Hogan:1986qda,Kosowsky:1991ua,Kosowsky:1992rz,Kosowsky:1992vn,Kamionkowski:1993fg}, has been widely studied in the literature. Such transitions can be readily realized in extended sectors in many realistic beyond the Standard Model (BSM) scenarios \cite{Schwaller:2015tja,Jaeckel:2016jlh,Dev:2016feu,Baldes:2017rcu,Tsumura:2017knk,Okada:2018xdh,Croon:2018erz,Baldes:2018emh,Prokopec:2018tnq,Bai:2018dxf,Breitbach:2018ddu, Fairbairn:2019xog, Helmboldt:2019pan,Ertas:2021xeh,Jinno:2022fom}. The dynamics of such bubbles admit various interesting phenomena, most notably the production of gravitational waves (GWs)\cite{Grojean:2006bp,Caprini:2015zlo,Caprini:2018mtu,Caprini:2019egz} (for a recent review see \cite{Athron:2023xlk}). 

The focus of this paper is the phenomenon of particle production during FOPTs. In the GW literature, particle production from FOPTs is generally considered in the context of interactions between the fast-moving bubble walls and the surrounding plasma \cite{Bodeker:2017cim,Hoche:2020ysm,Azatov:2020ufh,Gouttenoire:2021kjv, Baldes:2023fsp, Ai:2023suz}, which is known to provide a source of friction that affects bubble dynamics and the subsequent spectrum of GWs. In this paper, we focus on particle production that occurs purely due to the dynamics of the background field itself, independent of the presence or nature of a thermal plasma. Note that such particle production, analogous to GW production, is expected to be a far stronger phenomenon, since particle coupling strengths are significantly larger than (Planck suppressed) gravitational couplings. If this particle production mechanism is particularly efficient, it can also affect the dynamics of bubble collisions and the subsequent production of GWs. 

There are only a handful of papers in the literature that study particle production from background field dynamics during FOPTs. The first work to study this phenomenon in detail, by Watkins and Widrow \cite{Watkins:1991zt}, explored classical scalar wave production from two-bubble collisions and direct quantum particle production from such processes. Subsequent work by Konstandin and Servant \cite{Konstandin:2011ds} performed more detailed studies of this phenomenon in the context of cold baryogenesis. Building on these results, the most detailed study of this process was performed in a paper by Falkowski and No \cite{Falkowski:2012fb}, which studied particle production at a first order electroweak phase transition, and the possibility for this to account for dark matter. Several physical aspects of the process have recently been clarified in \cite{Shakya:2023kjf}. Ref.\,\cite{Katz:2016adq} discussed the issue of gauge invariance of the operators in a gauged theory. Most recently, Ref.\,\cite{Giudice:2024tcp} demonstrated that the formalism used to calculate particle production with this formalism is gauge-dependent, but it is nevertheless possible to extract results that are physically relevant. These papers found that bubble collisions during FOPTs could be a  viable source of heavy particles far above the temperature of the thermal bath, since the energy density in the boosted bubble walls at collision can be several orders of magnitude higher than the background scalar field vacuum expectation value (vev) or the ambient bath temperature. Such possibilities could be of particular interest for various beyond the Standard Model (BSM) applications, as has been explored, e.g.\,, in \cite{Falkowski:2012fb,Katz:2016adq,Freese:2023fcr, Giudice:2024tcp, Cataldi:2024pgt}.

Most of the above studies relied on semi-analytic estimates of particle production in two simplified settings: perfectly elastic collisions between bubbles, where the bubble walls simply reflect off each other without any further dynamics, or totally inelastic collisions, where the bubble walls stick together upon collision and dissipate all of their energy into scalar waves. The purpose of this paper is to perform numerical studies of bubble collision processes in more realistic scenarios that capture crucial physical effects missed in these studies, and provide greater physical insights into the process. 

In Section \ref{sec:formalism} we introduce the formalism to track the dynamics of the scalar field after bubble collisions and calculate particle production from such dynamics, and describe our numerical setup. In Section \ref{sec:results}, we present our results, highlighting the main physical aspects and crucial differences from existing results, and provide easy to use fit formulae corresponding to our numerical results for general use.
Section \ref{sec:discussion} summarizes our main results and discusses some broader aspects.

\section{Formalism}
\label{sec:formalism}

In this section, we present the formalism necessary to calculate particle production from the collision of two bubbles, and describe the details of our numerical setup. For simplicity, we work in the thin wall approximation, and consider relativistic bubble walls traveling at approximately the speed of light $v_w\approx 1$. We study the collisions in (1+1) dimensions, which are nevertheless applicable to 3D setups where the bubbles are sufficiently large at collision that the walls can be assumed to be planar at particle physics scales, which is generally a reasonable approximation.  

\subsection{Scalar Field Dynamics}
\label{subsec:dynamics}

\begin{figure}
    \centering
     \!\!\!\!\begin{subfigure}{0.5 \textwidth}
    \includegraphics[width= 1.05\textwidth]{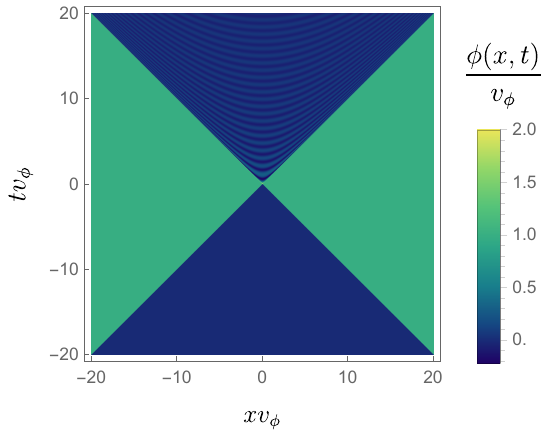}
     \caption{Elastic collision ($\epsilon=0.6, \; a=28.4$)}
    \end{subfigure}
    \begin{subfigure}{0.5 \textwidth}
    \includegraphics[width= 1.05\textwidth]{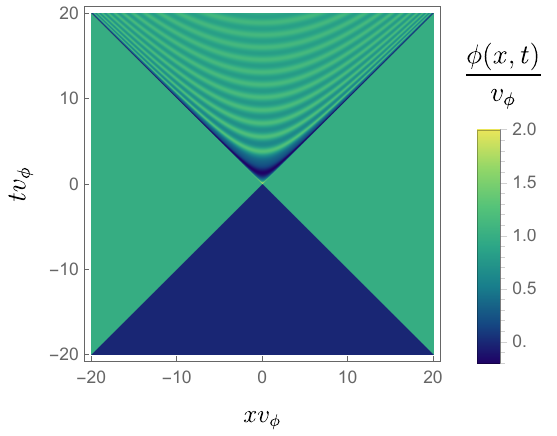}
    \caption{inelastic collision ($\epsilon=0.1, \; a=4.4$)}
    \end{subfigure}
   
    \caption{Scalar field spacetime configurations corresponding to elastic and inelastic bubble collisions (collisions occur at the origin).} 
    \label{fig:field-cnfs}
\end{figure}

The background scalar field configuration before collision is straightforward: it consists of two bubble walls approaching each other, with the region in between the two walls in the false vacuum and the region behind the walls (on the other sides) in the true vacuum.\,\footnote{For thick walled bubbles, the scalar field might not be at its true minimum anywhere inside the bubble, and instead evolves towards the true minimum and performs oscillations around it as the bubble expands (see e.g.\,\cite{Cutting:2020nla}). We do not consider such setups in this paper.} When the bubble walls collide, the field undergoes a local excitation at the collision point. 
In relativistic collisions, the bubble walls superpose and the field value at the point of collision is given by (see \cite{Jinno_2019})
\be
   \phi_{\mathrm{after}}=2\phi_{\mathrm{T}}-\phi_{\mathrm{F}},
    \label{eq.initialkick}
\ee
where $\phi_{\mathrm{T}}$ and $\phi_{\mathrm{F}}$ denote the field values inside and outside the bubbles, corresponding to the true and false vacua, respectively. Since $\phi_{\mathrm{after}}$ generally does not correspond to a local minimum of the potential, the field classically rolls down the potential and oscillates around one of the minima.\,\footnote{An exception is a periodic potential, for which $\phi_{\mathrm{after}}$ is also a minimum, hence the new field value after collision is stable and no oscillations occur after the collision \cite{Watkins:1991zt,Konstandin:2011ds,Falkowski:2012fb}.} Depending on the details of the potential, there are two possible scenarios for the subsequent evolution (see \cite{Watkins:1991zt,Konstandin:2011ds,Falkowski:2012fb} for more details; also see Fig.\,\ref{fig:field-cnfs} for illustrations of the two cases):
\begin{itemize}
\item \textit{Elastic collisions:} In such collisions, which occur when the minima are (almost) degenerate, the two colliding bubble walls reflect off each other and the false vacuum is re-established in the region between the receding walls: this corresponds to the scalar field climbing back over the barrier and into the false vacuum. 
Vacuum pressure eventually drives the walls back to each other, and the walls collide again; this process is repeated several times, with the walls becoming less energetic with each collision, before the true vacuum is finally established. 

\item \textit{Inelastic collisions:} In such scenarios, which occur for potentials with a relatively shallow barrier, there is no re-establishment of the false vacuum (that is, the field settles in the true vacuum region); instead, the collision converts the bubble wall energy to localized field oscillations around the true vacuum.  

\end{itemize}

To study the behavior of the scalar field in generic collision scenarios, we make use of the so-called trapping equation from Ref.\,\cite{Jinno_2019}, which allows for a simple analytical description of the field configuration in the relativistic bubble wall limit $v_{\mathrm{w}}\approx 1$, and can determine whether the collision is elastic or inelastic. The equation of motion of the field after the collision can be written in terms of a single variable \cite{Jinno_2019}
\begin{equation}
    \partial_{\mathrm{s}}^2 \phi + \frac{1}{s}\partial_{\mathrm{s}} \phi + \frac{\mathrm{d}V(\phi)}{\mathrm{d}\phi}=0,\,
    \label{eq:trapping}
\end{equation}
where $s$ is the light-front coordinate $s=\sqrt{t^2-x^2}$, and $V(\phi)$ is the scalar potential. The field value at the collision point after the collision is given by \Eref{eq.initialkick}, $\phi_{\mathrm{after}}=2\phi_{\mathrm{T}}-\phi_{\mathrm{F}}$, with vanishing velocity, which provides the initial conditions for solving this trapping equation.
  
\begin{figure}
    \centering
    \begin{subfigure}{0.49 \textwidth}
    \includegraphics[width= 0.9\textwidth]{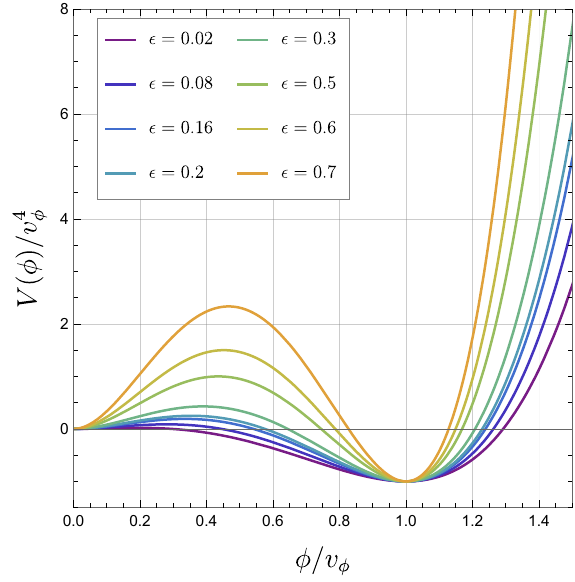}
    \end{subfigure}
    \begin{subfigure}{0.49 \textwidth}
    \includegraphics[width= 0.9\textwidth]{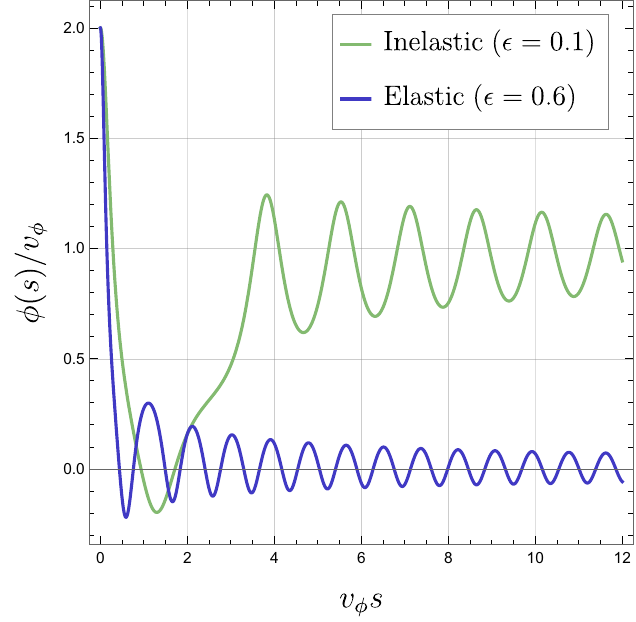}
    \end{subfigure}
    \caption{Left panel: Shape of potential for various $\epsilon$ values. Right Panel: Scalar field evolution after collision for representative elastic and inelastic cases.}
    \label{fig:shapeosc}
\end{figure}

Following \cite{Jinno_2019}, we consider the following generic potential for the scalar field $\phi$:
        \begin{equation}
		V(\phi)=a v_{\phi}^2 \phi^2 - (2a+4) v_{\phi} \phi^3+ (a+3)\phi^4.
        \label{eq:toypotential}
	\end{equation}
This potential has a metastable minimum (false vacuum) at $\phi=0$ and a global minimum (true vacuum) at $\phi=v_{\phi}$. One can define a degeneracy parameter $\epsilon$, determined by the dimensionless parameter $a$, as
 \begin{equation}
     \epsilon \equiv \frac{V_{max}-V(\phi=0)}{V_{max}-V(\phi=v_{\phi}) }= \frac{a^3 (a+4)}{a^3(a+4)+16(a+3)^3}\,,
 \end{equation}
   where $V_{max}$ is the height of the potential barrier separating the two minima. The shape of the potential can be altered by changing the value of $a$ (equivalently $\epsilon$), as shown in \Fref{fig:shapeosc} (left panel): increasing $\epsilon$ increases the height of the barrier separating the two minima, and also makes the potential steeper at higher field values. Ref.\,\cite{Jinno_2019} finds that
 \beq
  ~~~~~\epsilon>0.22: \text{elastic collisions}, ~~~~~ \epsilon<0.22: \text{inelastic collisions}. 
  \eeq
In both elastic and inelastic collisions, the field gets excited to a value $\phi_{\mathrm{after}}=2 v_\phi$ beyond the true minimum, then rolls back over the barrier into the false vacuum regime. A shallow barrier (corresponding to small $\epsilon$) enables the field to roll $\textit{again}$ over the barrier and oscillate around the true minimum, corresponding to an inelastic collision. In contrast, a higher barrier (corresponding to large $\epsilon$) $\textit{prevents}$ the field from rolling back over the barrier, trapping it instead in oscillations around the false vacuum, resulting in elastic collisions where the bubble walls do not disappear but bounce back, re-establishing the false vacuum in the region in between.  In both cases, the collision is therefore followed by scalar field oscillations around the corresponding minimum. 

In \Fref{fig:shapeosc} (right panel), we plot the evolution of the scalar field corresponding to these two cases, obtained by solving the trapping equation \Eref{eq:trapping}, where oscillations around the two vacua ($\phi=0\,(v_\phi)$ for false (true) vacuum) are clearly visible. In Fig.\,\ref{fig:field-cnfs}, we plot the spacetime configurations of the scalar field for these two cases. We show the two incoming relativistic bubble walls (approximated as step functions), which collide at the origin, leading to scalar field oscillations in the true and false vacua in the elastic and inelastic cases, respectively.  Realistically, the elastic case should feature multiple collisions between the bubble walls. However, the trapping equation can only track the evolution of the field in the region between the walls but cannot incorporate these repeated collisions. In this paper, we therefore only consider a single collision between the walls for the elastic as well as inelastic cases, which will nevertheless capture the main physical effects relevant for particle production that we are interested in. Ref.\cite{Watkins:1991zt} showed, through numerical simulations, that elastic collisions lose roughly $30\% - 40\%$ of the initial wall energy at every collision over a wide range of parameters. By energy conservation arguments, this implies that the boost factor at subsequent collisions is reduced by $30\% - 40\%$, and the typical separation distance between the walls is $60 - 70\%$ of the initial distance. The numerical simulations in \cite{Watkins:1991zt}  also show that most of the energy is radiated away after a few collisions. This occurs because when $30\% - 40\%$ of the energy is lost at each collision, after a few collisions the scalar field does not have energy to climb over the barrier into the false vacuum again, ending the elastic collision phase. Consequently, we can expect the particle number densities calculated here to be enhanced by an $\mathcal{O}(1)$ factor when multiple elastic collisions are taken into account.

The dominant oscillation frequency in either case is determined by the effective mass in the relevant vacuum; for the toy potential above, these masses are
    \be
      m_f^2 = \frac{d^2 V(\phi)}{d \phi^2} \rvert_{\phi=0}= 2a v_{\phi}^2, ~~~~~~
      m_t^2 = \frac{d^2 V(\phi)}{d \phi^2} \rvert_{\phi= v_{\phi}}= 2 v_{\phi}^2 (a+6). 
      \label{eq:masses}
  \ee

 \subsection{Mode Decomposition and Particle Production}

 We now describe the method to calculate particle production using the scalar field configurations obtained in the previous subsection. We will follow the formalism introduced in \cite{Watkins:1991zt} and further developed in \cite{Konstandin:2011ds, Falkowski:2012fb}; the interested reader is referred to these papers for further details. 

The formalism consists of treating the moving and colliding bubble walls and subsequent oscillations as classical external field configurations of the scalar $\phi (x,t)$. The probability for this configuration to decay into particles is extracted from the imaginary part of its effective action, which is evaluated by decomposing $\phi (x,t)$ via a Fourier transform into modes of definite frequency $\omega$ and momenta $k$. Modes of four-momentum $\chi=\omega^2-k^2 >0$ can be interpreted as off-shell propagating field quanta of $\phi$ with mass $m^2=\chi$, and the probability for each such mode excitation to decay is given by the imaginary part of its 2-point 1PI (1 particle irreducible) Green function. 

Assuming each decay produces a pair of identical particles, and assuming planar walls in 3-dimensional space, the number of particles produced per unit area of bubble wall can be expressed as \cite{Watkins:1991zt,Konstandin:2011ds, Falkowski:2012fb}
\be
\frac{N}{A}= 2 \int\frac{dk\,d\omega}{(2\pi)^2}\,|\tilde{\phi}(k,\omega)|^2 \,\text{Im}[\tilde{\Gamma}^{(2)}(\omega^2-k^2)]\,.
\label{particle}
\ee
Here $\tilde{\phi}(k,\omega)$ is the Fourier transform of the field configuration $\phi(x,t)$.
From the Optical Theorem, the imaginary part of the 2-point 1PI Green function is given by the sum over matrix elements of all possible decay processes:
\be
\text{Im} [\tilde{\Gamma}^{(2)}(\chi)]=\frac{1}{2}\sum_\alpha \int d\Pi_\alpha |\bar{\mathcal{M}}(\phi\to\alpha)|^2\,\Theta(\chi-\chi_{min\,(\alpha)}).
\label{optical}
\ee
Here the sum runs over all possible final states $\alpha$ that can be produced, $|\bar{\mathcal{M}}(\phi\to\alpha)|^2$ is the spin-averaged squared amplitude for the decay of $\phi$ into the given final state $\alpha$, $d\Pi_\alpha$ is the relativistically invariant n-body phase space element, $\Theta$ is the Heaviside step function, and $\chi_{min\,(\alpha)}=(\sum m_\alpha)^2$ represents the minimum energy required to produce the final state particles on-shell. For $n$-body final states, one should replace the prefactor $2$ in \Eref{particle} by the appropriate number. It should be noted that the evaluation of $|\bar{\mathcal{M}}(\phi\to\alpha)|^2$ can in principle be gauge dependent since the formalism considers off-shell excitations of the field configuration, requiring careful consideration for the production of gauge bosons in particular \cite{Giudice:2024tcp}.

Using $\chi=\omega^2-k^2$ and $\xi=\omega^2+k^2$, and integrating over $\xi$, the number and energy of particles produced per unit area can be simplified to  \cite{Falkowski:2012fb}
\bea
\frac{N}{A}&=&\frac{1}{4\pi^2}\int_{\chi_{min}}^{\chi_{max}} d\chi\,f(\chi) \,\text{Im} [\tilde{\Gamma}^{(2)}(\chi)],\nonumber\\
\frac{E}{A}&=&\frac{1}{8\pi^2}\int_{\chi_{min}}^{\chi_{max}} d\chi\,f(\chi)\,\sqrt{\chi}\, \text{Im} [\tilde{\Gamma}^{(2)}(\chi)]\,.
\label{numberenergy}
\eea

Here $f(\chi)$ encapsulates the relevant details of the underlying field configuration in the Fourier decomposition and hence represents the \textit{efficiency factor} for particle production at the given scale $\sqrt{\chi}$. The lower limit $\chi_{min}$ is set either by the masses of the particle species being produced or by the physical infrared (IR) cutoff scale of the configuration, which corresponds to the maximal bubble size $R_*$ at collision. The upper, ultraviolet (UV) cutoff is given by $\chi_{max}=(\gamma_w/l_w)^2$, the boosted bubble wall thickness, which represents the maximum energy scale probed by the process. Note that \Eref{numberenergy} correspond to the case when each excitation decays into two particles, and should be appropriately modified for multi-body final states.  

From \Eref{particle} and \Eref{numberenergy}, it is clear that the computation consists of two independent components. The first consists in determining the efficiency factor $f(\chi)$ from the Fourier transform $\tilde{\phi}(k,\omega)$ of the field configuration; this depends purely on the spacetime dynamics of the background scalar field during the phase transition. The second piece corresponds to the ``particle physics" aspect of the setup: calculating the decay probabilities of the excitations via the computation of $|\bar{\mathcal{M}}(\phi\to\alpha)|$ for all possible decays in a given model. The focus of this paper will be on the calculation of $f(\chi)$ for various realistic field configurations.

\subsection{Numerical Setup}

In this subsection, we describe the details of our numerical setup. We study the scalar field over a (1+1) dimensional spacetime region over the interval $[-L/2,L/2]$ for both $x$ and $t$, with bubble collision taking place at the origin $(0,0)$ (see Fig. \ref{fig:field-cnfs}). The $t<0$ region thus consists of two bubble walls approaching each other; for simplicity, we describe this part by two step function walls approaching each other at the speed of light $v_w=1$, with $\phi=0$ in the region between them, and $\phi=v_\phi$ on the other sides. In principle, the walls have some finite thickness $l_w$; however, at large boosts $\gamma_w$, this thickness is Lorentz contracted and cannot be resolved in the numerical studies, hence we simply treat them as step functions. The field configuration after collision (in the $t>0$ region) is obtained by solving the trapping equation (\Eref{eq:trapping}) for a given scalar potential (as parameterized by \Eref{eq:toypotential}).

We generate $N \times N$ arrays describing the field configuration $\phi(x,t)$ in this finite $x-t$ plane; thus, the field is sampled at spatial and temporal intervals $d=L/N$. We then numerically perform a discrete Fourier transform of this configuration to obtain the efficiency factor $f(\chi)$ as described in the previous subsection. 

There are three physical scales of relevance: the bubble size at collision $R_*$ (typically a few orders of magnitude smaller than the inverse Hubble size $H^{-1}$ at the time of the transition), the scale of scalar field oscillations after collision, which is given by the mass of the field in the relevant vacuum, $m_{t,f}\sim \mathcal{O}(v_\phi)$ (see \Eref{eq:masses}), and the boosted wall thickness scale $l_w/\gamma_w$. Typically, the three scales are very different, $R_*\gg v_\phi^{-1} \gg l_w/\gamma_w$, generally separated by several orders of magnitude. Realistically, one ought to choose $L\gtrsim R_*$, and choose spatial and temporal resolutions such that the other two scales are resolved. In practice, this is computationally very challenging, as is well known to the gravitational waves simulation community. Since the primary goal of our study is to capture particle production effects, the relevant scale of interest is the particle physics scale $v_\phi$, and we choose values of $L$ that are a few orders of magnitude larger than $v_\phi^{-1}$. Unless stated otherwise, the default choices for the spatial/temporal size and resolution (sampling intervals) for our numerical studies are
\begin{equation}
    L= 40 \, v_{\phi}^{-1} \qquad \mathrm{and} \qquad d=0.01 \, v_{\phi}^{-1}\,.
    \label{eq:choices}
\end{equation}
In all of our numerical studies, we ensure $d< 0.1 \, m_{t,f}^{-1}$, so that the scalar field oscillations are always properly resolved despite the finite sampling.
Note that we cannot resolve the boosted bubble wall thickness for large $\gamma_w$ (hence the step function approximation), and can only study a small fraction of the full bubble size.  Nevertheless, the above setup is sufficient to capture all the important details of particle production.  

Note that the finite size and resolution of our numerical studies incur various limitations. For the choices in \Eref{eq:choices}, our studies can only resolve $\omega, k$ between $0.05\,\pi\,v_\phi$ and $100\,\pi\,v_\phi$. In practice, the results can also deteriorate and give spurious features at scales close to these two limits. Indeed, in our numerical studies we do observe spurious effects close to the lower momentum cutoff (i.e.\,low $k$ values), which manifest as imaginary components of the Fourier transform. In our calculations and results below, we therefore take only the real part of the Fourier transform, which proves to be effective in eliminating such spurious unphysical contributions (see Appendix \ref{app:imaginary} for a more detailed discussion).

\section{Results}
\label{sec:results}

In this section, we present the results of our numerical study. We first discuss our results for the efficiency factors, comparing them with existing results in the literature (from \cite{Watkins:1991zt,Konstandin:2011ds,Falkowski:2012fb}), before turning to calculations of particle production.  

\subsection{Mode Decomposition and Efficiency Factors}
\label{subsec:mode}

\begin{figure}[t]
    \centering
    \!\!\!\!\begin{subfigure}{0.5 \textwidth}
   \includegraphics[width= .95\textwidth]{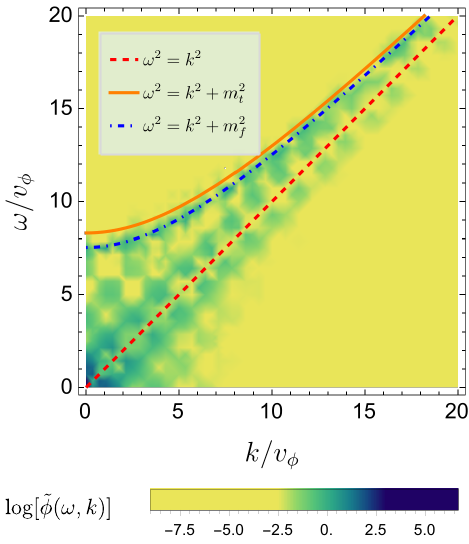}
     \caption{elastic collision ($\epsilon=0.6, \; a=28.4$)}
    \end{subfigure}
    \begin{subfigure}{0.5 \textwidth}
     \includegraphics[width= 0.95\textwidth]{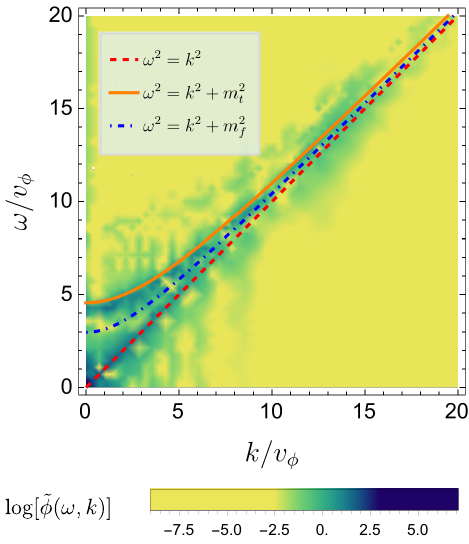}
     \caption{inelastic collision ($\epsilon=0.1, \; a=4.4$)}
    \end{subfigure}
    \caption{Density plots of the logarithm of the Fourier transforms of field configurations describing elastic (left panel) and inelastic (right panel) collisions. We also show contours corresponding to $\omega^2=k^2$ and $\omega^2=k^2+m_{t,f}^2$ for effective mass $m_{t}\,(m_f)$ in the true (false) vacua. }
    \label{fig:FTs}
\end{figure}

In \Fref{fig:FTs}, we plot the Fourier transform in the $\omega-k$ plane for representative elastic and inelastic collision configurations (corresponding to those shown earlier in \Fref{fig:shapeosc}(b) and \Fref{fig:field-cnfs}). In both cases, the main contributions are seen to be clustered around two main branches, originating from distinct physical phenomena (see \cite{Shakya:2023kjf} for related discussions): 
\begin{itemize}
\item around $\omega^2=k^2:$ these correspond to excitations induced by the process of collision between the two bubble walls. 
\item around $\omega^2=k^2+m^2:$ these correspond to excitations induced by the scalar field oscillating around a minimum after the walls have collided, with $m$ representing the effective mass of the scalar around the corresponding minimum (false minimum for elastic collisions, true minimum for inelastic collisions) after the bubbles collide.  
\end{itemize}
These generic features remain applicable for other choices of $\epsilon$.

From this Fourier transform, we can compute the efficiency factor $f(\chi)$ as defined in \Eref{numberenergy}. This can be done for an arbitrary field configuration as follows.
Using $\xi=\omega^2+k^2$ and  $\chi=\omega^2-k^2$, we can write 
\begin{align}
    \frac{N}{A}&= \frac{1}{(2\pi)^2}\int_{-\gamma /l}^{\gamma /l} dk\int_{-\gamma /l}^{\gamma /l} d\omega \, |\tilde{\phi}(k,\omega)|^2 \,\text{Im}[\tilde{\Gamma}^{(2)}(\omega^2-k^2)]\nonumber\\
    &=\frac{4}{(2\pi)^2}\int_0^{\gamma /l} d k \int_0^{\gamma /l} d\omega \,|\tilde{\phi}(k,\omega)|^2 \,\text{Im}[\tilde{\Gamma}^{(2)}(\omega^2-k^2)] \nonumber\\
    &=\frac{4}{(2\pi)^2} \int_0^{(\gamma /l)^2} d \chi \,\text{Im}[\tilde{\Gamma}^{(2)}(\chi)] \int_{\chi}^{2(\gamma /l)^2-\chi} d\xi  \frac{1}{4 \sqrt{\xi^2-\chi^2}}  |\tilde{\phi}(\chi,\xi)|^2 \nonumber\\
    &= \frac{1}{(2\pi)^2}\int_0^{(\gamma /l)^2} d\chi \, \text{Im}  [\tilde{\Gamma}^{(2)}(\chi)] \,f(\chi)
    \label{def.Fchi}
\end{align}
Thus, for any function $\tilde{\phi}(k, \omega)$, the second integral in second-to-last step can be numerically evaluated for discrete values of $\chi$ to numerically obtain $f(\chi)$.

\begin{figure}
    \centering
  \!\!\!\!  \begin{subfigure}{0.5\textwidth}
       \includegraphics[width=0.95\textwidth]{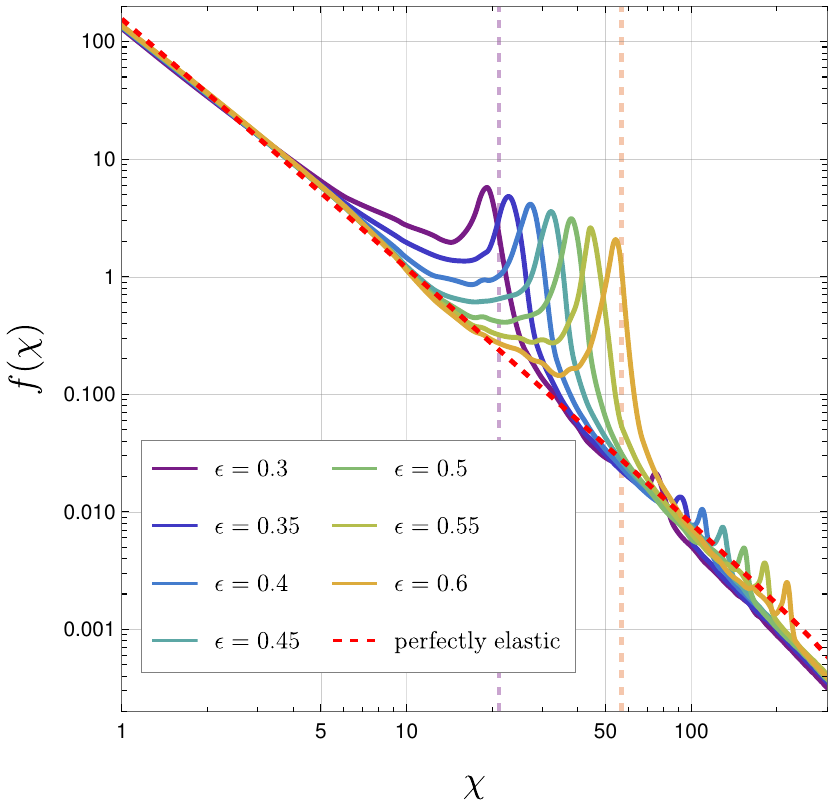}
       \caption{Efficiency factor for elastic collisions.}
    \end{subfigure}
      \begin{subfigure}{0.5 \textwidth}
      \includegraphics[width=0.95\textwidth]{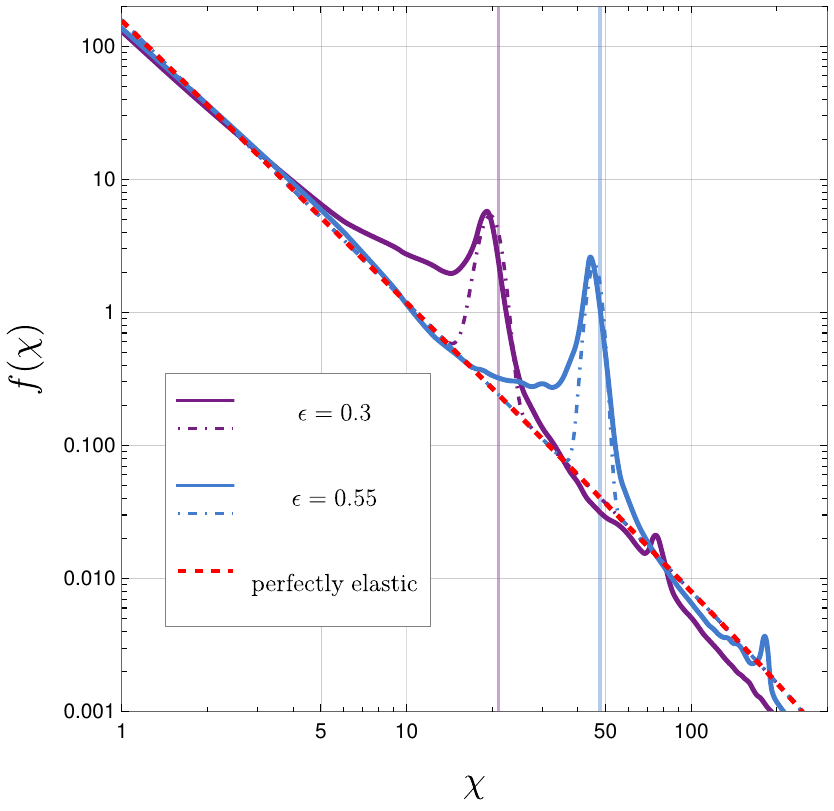}  
      \caption{Comparison with analytical results and fit functions.}
    \end{subfigure}\\
    \vskip 1cm
    \!\!\!\!  \begin{subfigure}{0.5 \textwidth}
      \includegraphics[width=0.95\textwidth]{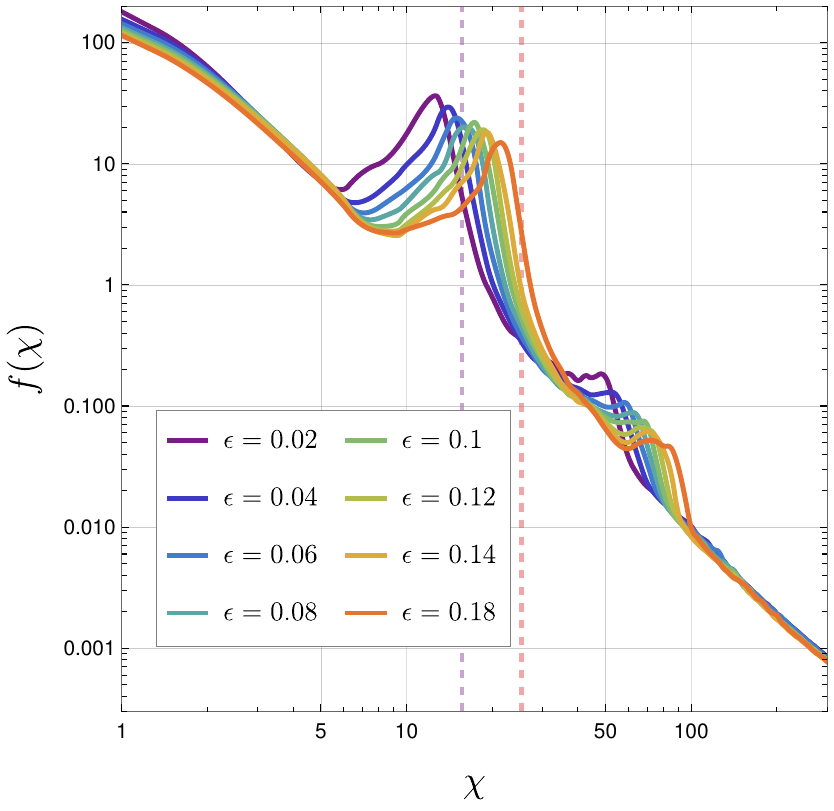}  
      \caption{Efficiency factor for inelastic collisions.}
    \end{subfigure}
      \begin{subfigure}{0.5 \textwidth}
      \includegraphics[width=0.95\textwidth]{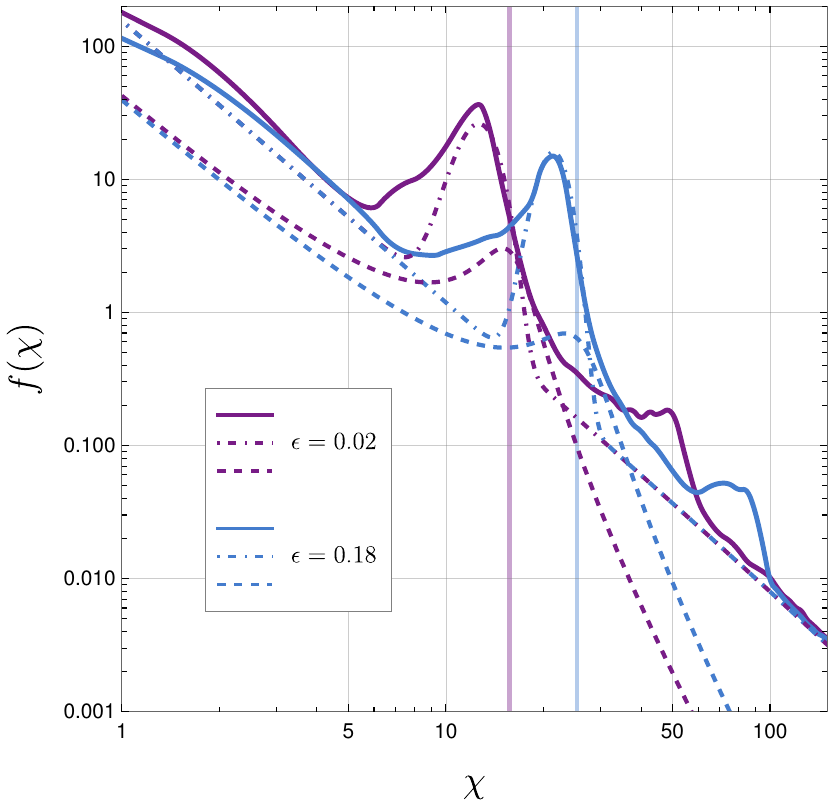}  
      \caption{Comparison with analytical results from previous literature (dashed) and our fit functions (dot-dashed).}
    \end{subfigure}
    \caption{Efficiency factor $f(\chi)$ in elastic (top panel) and inelastic (bottom panel) collisions for different potential barrier heights (as determined by $\epsilon$) as a function of $\chi$ (both axes are in units of $v_\phi$). To guide the eye, we also show the effective scalar mass for the highest and lowest values as dashed vertical lines in each plot on the left panels. For the elastic case, we also show (red dashed line) the analytic result for a perfectly elastic collision (\Eref{eq:fchielastic}). On the right panel, for comparison, we show the analytical results from Ref.\,\cite{Falkowski:2012fb} as dashed curves, and our fit functions \Eref{eq:elasticfit} and \Eref{eq:inelasticfit} as dot-dashed curves, for a few cases. For this plot, we have fixed $\gamma_w=200$ and $l_w=10 v_{\phi}^{-1}.$
}
    \label{fig:efficiencyfactors}
\end{figure}

 In \Fref{fig:efficiencyfactors},  we plot $f(\chi)$ for various choices of $\epsilon$ (which correspond to various barrier heights in the scalar potential) for elastic (top panel) and inelastic (bottom panel) collisions in the left column. In the panels in the right column we also plot the comparisons with analytic results as calculated in \cite{Falkowski:2012fb} (dashed curves), as well as our fit functions from \Eref{eq:elasticfit} and \Eref{eq:inelasticfit} (dot-dashed curves), which we will describe in greater detail below. 

For a perfectly elastic collision, i.e. when the walls bounce back with the same relative speed and there is no energy dissipation in scalar waves, the efficiency factor can be calculated analytically as \cite{Falkowski:2012fb} 
\be
f_{\mathrm{PE}}(\chi)=\frac{16 v_{\phi}^2}{\chi^2}\, \text{Log}\left[\frac{2(\gamma_w/l_w)^2-\chi+2(\gamma_w/l_w)\sqrt{(\gamma_w/l_w)^2-\chi}}{\chi}\right]\,\Theta[(\gamma_w/l_w)^2-\chi]
\label{eq:fchielastic}
\ee
Thus the function scales approximately (up to the logarithmic factor) as $\sim \chi^{-2}$ (this is also consistent with results in \cite{Watkins:1991zt,Konstandin:2011ds}). This function is plotted as the red dashed curve in the top panels of \Fref{fig:efficiencyfactors}.  

For both elastic and inelastic collisions, the efficiency factors obtained from our numerical studies are seen to approximately follow the power law $\propto \chi^{-2}$  consistent with the analytical result for a perfectly elastic collision \Eref{eq:fchielastic}; this component originates from the collision between the two bubble walls, which is a very rapid process and thus contains very high frequency components. In addition, the numerical results show a prominent bump in the spectrum, peaking at $\chi\approx m_{\mathrm{f}}^2$ for elastic collisions and at $\chi\approx m_{\mathrm{t}}^2$ for inelastic collisions; this feature can be attributed to the field oscillations around the corresponding minimum after the bubble walls collide and the scalar field gets excited away from the minima. We expect these two features -- a power law $\propto \chi^{-2}$ and a bump peaking at the scalar mass in the vacuum that is realized after collision -- to coexist for any realistic potential. Different values of $\epsilon$ correspond to different shapes of the potential, hence different masses in the two vacua, and we see the positions of the peaks shifting accordingly in both plots. The vertical dashed lines in the various panels show the relevant mass scales for a few cases.

Let us briefly discuss the main differences between our numerical results and the analytic solutions from \cite{Falkowski:2012fb}, as seen in the right panels of \Fref{fig:efficiencyfactors}. For elastic collisions (top row), the analytic limit \Eref{eq:fchielastic} misses the bump around $\chi=m_f^2$ from scalar field oscillations after the collision we see for realistic elastic collisions, but otherwise agrees well with our numerical results. As discussed earlier, the oscillations are inevitable for generic (non-periodic) potentials, and therefore represent an important feature that is missed by the simplified analytic treatment. There are
small discrepancies from the $\chi^{-2}$ scaling at $\chi > m_f^2$, likely a combination of energy getting lost to scalar oscillations and numerical effects from the limitations of
the numerical study, but the results continue to agree up to $\mathcal{O}(1)$ factors.

For inelastic collisions, we show (dashed curves in panel (d)) the analytic result from  \cite{Falkowski:2012fb} for a totally inelastic collision
\begin{equation}
    f_{\mathrm{TI}}(\chi)= \frac{4 v_{\phi}^2 m_{t}^4 }{\chi^2 \left[ (\chi-m_{t}^2)^2+m_{t}^6\left(\frac{l_w}{\gamma_w}\right)^2 \right]} \mathrm{Log} \left[ \frac{2 \left(\frac{\gamma_w}{l_w}\right)^2 + \chi + 2\frac{\gamma_w}{l_w} \sqrt{\left(\frac{\gamma_w}{l_w}\right)^2+ \chi}}{\chi}\right].
    \label{eq:fchiTI}
\end{equation}
This result was derived for a quadratic approximation of the potential well around the true vacuum, and solving for the field evolution after the collision (see \cite{Falkowski:2012fb} for details). There are two notable features: (1) the efficiency factor features a peak around $\chi=m_t^2$, corresponding to scalar field oscillations after collision, and (2) at momenta higher than the mass, the value drops off as $f(\chi)\sim \chi^{-4}$, deviating from the $\chi^{-2}$ behavior seen in our numerical results as well as the perfectly elastic limit. The steeper power law scaling can be understood by noting that the analytic result in \cite{Falkowski:2012fb} is only derived for the field configuration after collision, and assuming that the energy in the walls is completely transferred to scalar field oscillations.  In particular, in \cite{Falkowski:2012fb} the Fourier transform for a perfectly inelastic collision is reported to be
\begin{equation}
    \Tilde{\phi}_{\mathrm{inelastic}}(k,\omega)= \frac{\pi l_{\mathrm{w}} k}{2 \gamma_{\mathrm{w}}} \frac{2 v_{\phi}}{\sinh\left(\frac{\pi l_{\mathrm{w}} k}{2 \gamma_{\mathrm{w}}}\right)}
    \left( 
    \frac{1}{\omega^2-k^2} - \frac{1}{\omega^2-k^2- m_{t}^2} \right),
    \label{eq:ftinelastic}
  \end{equation}
which vanishes in the limit $m_{t}\to 0$, i.e. the potential becomes flat and there are no oscillations. Therefore, this analytic approach misses the contribution from the collision process before the scalar waves develop. As our numerical results demonstrate, considering the spacetime region including the entire collision process restores the $f(\chi)\sim \chi^{-2}$ scaling at high momenta.
 Therefore, the analytic results from \cite{Falkowski:2012fb} capture the peak in the efficiency factor from oscillations but severely underestimate the contribution at $\chi>m_t^2$. Note, however, that the peak heights derived from the analytical result and numerical studies are markedly different; this discrepancy will be explored in detail in the next subsection.

Finally, note that the numerical results contain secondary peaks at higher $\chi$, as can be seen in all panels of \Fref{fig:efficiencyfactors} -- these are very prominent for inelastic collisions, but also exist for elastic collisions.  In all cases, these appear at $\chi\approx(2m_{\mathrm{f,t}})^2$. It is unclear to us whether these higher harmonics have a physical origin, or are simply artifacts of the numerical procedure, arising from taking the Fourier transform over a finite spacetime interval.  In any case, they generally contribute negligibly to calculations of particle number densities, hence we ignore these secondary peaks in our subsequent discussions.

\subsection{Characterization of the Peak}
\label{subsec:peak}

\begin{figure}
    \centering
    \begin{subfigure}{0.49 \textwidth}     
    \includegraphics[width=1 \textwidth]{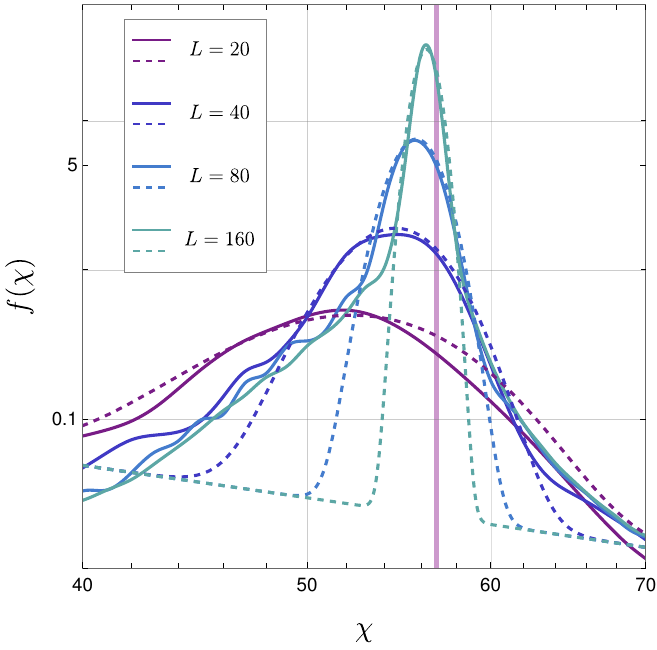}
    \caption{elastic cases ($\epsilon=0.6$)}
    \end{subfigure}
     \begin{subfigure}{0.49 \textwidth}     
    \includegraphics[width=1 \textwidth]{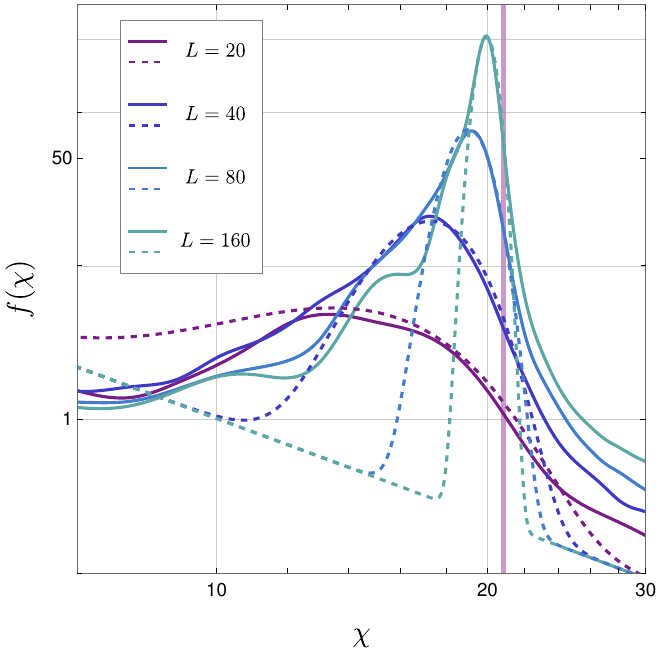}
    \caption{inelastic cases ($\epsilon=0.1$)}
    \end{subfigure}
    \caption{Efficiency factor peak dependence on the size of the field configuration considered (parameterized by $L$, shown here in units of $v_\phi$) in elastic (left panel) and inelastic (right panel) cases (axes in units of $v_\phi$). The dashed curves are obtained using the fit functions in  \Eref{eq:elasticfit} and \Eref{eq:inelasticfit}. The solid vertical lines correspond to the effective mass of the scalar field in the relevant vacuum. All dimensionful quantities are in units of $v_\phi$.
    }
    \label{fig:peaks}
\end{figure}

Beyond the power law feature, the peak due to oscillations is the most important feature in the efficiency factor spectrum (\Fref{fig:efficiencyfactors}). We now examine its properties in greater detail. A peak is characterized by three major properties: position, width, and amplitude, which are very different between our numerical results and the analytic results from \cite{Falkowski:2012fb} (as seen, e.g. in \Fref{fig:efficiencyfactors}(d))). For our results, we find that all three properties are sensitive to the size $(L)$ of the spacetime region chosen; we plot the peaks for different choices of $L$ for representative elastic and inelastic collision scenarios in \Fref{fig:peaks}.

For the analytic solution from \cite{Falkowski:2012fb}, given by \Eref{eq:fchiTI}, the peak always occurs at the scalar mass in the true vacuum, corresponding to oscillations around the true minimum. In contrast, our numerical results in \Fref{fig:efficiencyfactors} and \Fref{fig:peaks} show that the peaks occur slightly below the mass of the scalar for both elastic and inelastic collisions. This behavior can be understood by noting (see e.g.\,right panel of \Fref {fig:shapeosc}) that the initial field excitations after the collision are not oscillations, and the oscillations in the first few cycles are not symmetric around the corresponding minimum, which distort the peak away from the true mass. As more and more ``proper" oscillations around the minimum are incorporated by including a larger number of oscillation cycles -- which occurs when a larger region of spacetime is considered, i.e.\, for larger values of $L$ -- the peaks indeed shift closer and closer to the true mass, as seen in \Fref{fig:peaks}. 

The peak widths also reflect this behavior. Because of these initial non-oscillatory and non-symmetric excitations, the peaks are quite broad, and neither smooth nor symmetric around the scalar mass. Again, as more ``proper" oscillations are incorporated for larger $L$ values, the peaks are seen to become smoother and more symmetric. Furthermore, for larger L, as the scalar waves spread out in space and the oscillation amplitude becomes smaller,  the quadratic approximation for the potential around the minimum becomes better and better, hence the peaks get more sharply pronounced around the scalar mass, as is clearly visible in \Fref{fig:peaks}.  

Next, we comment on the amplitude of the peaks. The initial analytic result derived in \cite{Falkowski:2012fb} for the totally inelastic case was divergent at $\chi^2\to m_t^2$, which is simple to understand: since the calculation corresponds to all of spacetime and contains an infinite number of oscillation cycles, the efficiency factor diverges at $\chi^2\to m_t^2$ as there is infinite power at this frequency. In the corrected formula from \cite{Falkowski:2012fb} shown in \Eref{eq:fchiTI}, this divergence is regulated through the prescription
\be
(\chi-m_t^2)^2 \to (\chi-m_t^2)^2 + m_t^6 l_w^2/\gamma_w^2\,, 
\label{eq:regulation}
\ee
where $l_w/\gamma_w$ is the boosted wall thickness. However, the physical motivation for this correction is unclear: in the post-collision phase in a perfectly inelastic collision, the boosted wall thickness before collision is irrelevant. In fact, the comparison  with our numerical results (see \Fref{fig:efficiencyfactors}\,(d)) suggests that the prescription in \Eref{eq:regulation} is too extreme: the numerical results suggest that the amplitude of the peaks should be significantly larger. Furthermore, from \Fref{fig:peaks}, we see that the peak amplitude scales as $L^2$, which is indeed the correct scaling for the number of oscillation cycles contained in an $L\times L$ region of spacetime.

\subsection{Fit Functions}

We now provide easy to use fit functions to our numerical results that can be used to calculate the particle production efficiency factor $f(\chi)$ for generic setups. For a given model, one would need to determine whether the collisions are elastic or inelastic, and have the following quantities at hand:
\be
\{v_\phi, m_t, m_f, \gamma_w,l_w, R_*,\Gamma\}\,
\ee
Here $\Gamma$ is the decay rate of the scalar undergoing the phase transition, $R_*$ is the typical bubble size at collision, and the remaining parameters are as defined earlier. Using these parameters, the efficiency factors can be formulated as
\be
f_{\text{elastic}}(\chi)= f_{\mathrm{PE}}(\chi)+\frac{v_{\phi}^2L_p^2}{15 m_{\mathrm{t}}^2}\exp{\left(\frac{-(\chi - m_{\mathrm{t}}^2+12 m_{\mathrm{t}}/L_p)^2}{440 \, m_{\mathrm{t}}^2 / L_p^2}\right)}\qquad \text{(elastic collisions)}
\label{eq:elasticfit}
\ee
\be
f_{\text{inelastic}}(\chi)= f_{\mathrm{PE}}(\chi)+\frac{v_{\phi}^2L_p^2}{4 m_{\mathrm{f}}^2}\exp{\left(\frac{-(\chi - m_{\mathrm{f}}^2+31 m_{\mathrm{f}}/L_p)^2}{650 \, m_{\mathrm{f}}^2 / L_p^2}\right)}\qquad \text{(inelastic collisions)}
\label{eq:inelasticfit}
\ee
Here, $f_{\mathrm{PE}}$ is the efficiency factor for a perfectly elastic collision, given in \Eref{eq:fchielastic}, and $L_p=\text{min}(R_*, \Gamma^{-1})$. The dot-dashed curves in \Fref{fig:efficiencyfactors} and the dashed curves in \Fref{fig:peaks} are made using these formulae, but taking $L_p=L$ as defined for the numerical setup. We have checked that these formulae provide good fits for a broader range of $\epsilon$ values beyond those shown in the figures (see Appendix \ref{app:fits} for more examples). 

For the power law component, we have simply used the analytic result $f_{\mathrm{PE}}$ for perfectly elastic collisions \Eref{eq:fchielastic}, which provides good fits to our numerical results from both elastic and inelastic collisions. There are small discrepancies at large and small $\chi$ (see \Fref{fig:efficiencyfactors} right panels), likely due to some energy getting lost to scalar oscillations and possible numerical effects from the limitations of the numerical study. Nevertheless, since these only lead to 
 O(1) deviations, and as this form is motivated from physical considerations \cite{Watkins:1991zt, Falkowski:2012fb, Shakya:2023kjf}, we choose to keep the analytic formulation $f_{\mathrm{PE}}$. 

 We find that the peaks are well approximated by Gaussian functions of the forms shown in 
 \Eref{eq:elasticfit} and \Eref{eq:inelasticfit}, with the parameterization and numerical factors chosen purely to best match the numerical results for a wide range of potential shapes. In particular, note that we have added an offset factor in the numerator of the exponential to account for the fact that peaks shift away from the scalar mass $m_{t,f}$ when smaller regions of spacetime are considered. From the plots, it is clear that the fits capture the tips of the peaks very well, but break down further away from the peak as the peak shapes become non-Gaussian and irregular; nevertheless, we have checked that this introduces at worst an O(1) factor discrepancy for particle production calculations for any choice of parameters.  

Recall from the discussion in the previous subsection that the properties of the peak in the spectrum depend strongly on $L$, the size of the spacetime region considered. While $L$ is a parameter for our numerical studies, in the fit formulae we have prescribed instead the physical counterpart $L_p=\text{min}(R_*, \Gamma^{-1})$. Realistically, our derived results are only valid for scalar waves within a single bubble, of size $R_*$ (this cutoff scale for the peak was also mentioned in \cite{Falkowski:2012fb}); beyond this, scalar waves from multiple bubbles interfere, which is not captured by our study, hence $R_*$ is the maximal size for which our results should be valid. However, if the scalar waves decay rapidly before they propagate such distances, as can occur if the scalar has efficient decay channels, then the correct size of spacetime region to consider where scalar oscillations are relevant is instead the decay length, $\Gamma^{-1}$. Our prescription of $L_p=\text{min}(R_*, \Gamma^{-1})$ is therefore intended to select the smaller of these two physical scales. 

Finally, it is important to understand the limits within which our fit functions are applicable. As discussed earlier, due to the finiteness of the chosen region of spacetime, our studies can only resolve $\omega, k$ between $0.05\,\pi\,v_\phi$ and $100\,\pi\,v_\phi$, and, strictly speaking, our numerical results are only valid within these values. However, from the plots and fit functions above it is clear that the efficiency factor simply scales as a power law $\sim \chi^{-2}$ beyond these limits, and our fit functions can simply be extrapolated in this manner until additional physical effects become relevant. In the ultraviolet (UV), the results are not applicable for distances smaller than the boosted wall thickness $l_w/\gamma_w$, as effects related to the finite width of the bubble wall, which we have not taken into account, become important;  therefore, $\chi_{\text{max}}=\gamma_w^2/l_w^2$ represents the UV cutoff for the fit functions. Likewise, the infrared (IR) cutoff is given by the inverse of the bubble wall radius $R_*^{-1}$; at physical scales larger than this, the existence of multiple bubbles should be taken into account, which will modify the efficiency factor.  

\subsection{Particle Production}
\label{particleprod}

As an application of the results derived in the previous sections, and to highlight the differences compared to existing results in the literature  \cite{Watkins:1991zt,Konstandin:2011ds,Falkowski:2012fb}, in this section, we calculate particle production for a simple scenario. 

Consider a scalar $\psi$ that couples to $\phi$ via the interaction term $\frac{1}{2}\lambda v_\phi \phi \psi^2$. For this simple case, 
the imaginary part of the 2-point 1PI Green's function is
\be
        \mathrm{Im}\left( \Tilde{\Gamma}^{(2)}(\chi)\right)_{\phi^*\to\psi\psi} = \frac{\lambda^2 v_{\phi}^2}{32 \pi} \sqrt{1-\frac{4 m_{\psi}^2}{\chi}}~   \Theta[\,\chi-(2 m_\psi)^2].
        \label{specific}
\ee
Here, $m_\psi^2=\lambda v_\phi^2+m_0^2$; $\psi$ gets a mass contribution from the phase transition (first term) in addition to its bare mass $m_0$ (second term).

\begin{figure}
    \centering
    \begin{subfigure}{0.48\textwidth}
     \centering
     \includegraphics[width=1 \textwidth]{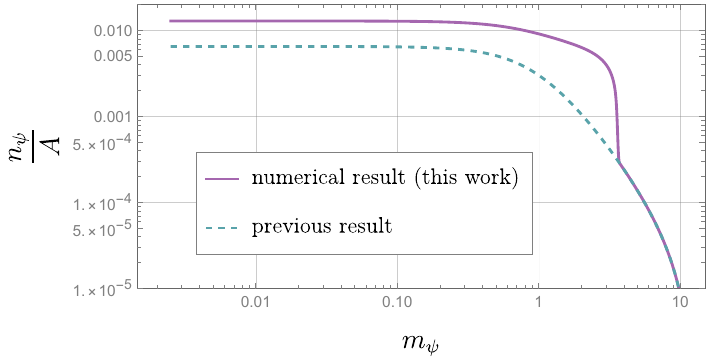} 
     \caption{Elastic collision ($\epsilon=0.6)$}
    \end{subfigure}  
    \begin{subfigure}{0.455\textwidth}
     \centering
     \includegraphics[width=1\textwidth]{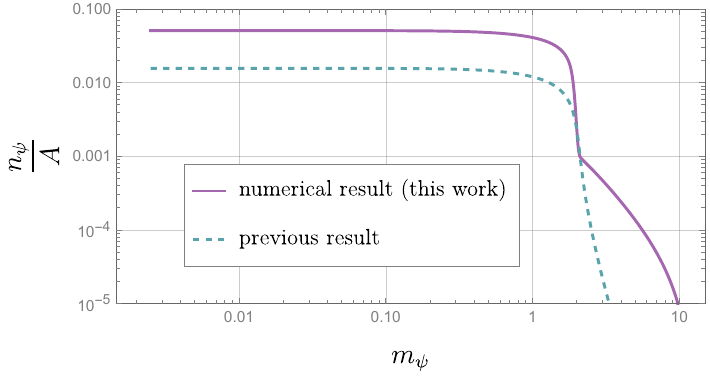} 
          \caption{Inelastic collision ($\epsilon=0.1)$}plots
    \end{subfigure} 
    \caption{ Particle number produced per unit bubble wall area as calculated using the fit functions to our numerical results (solid curves) and from previous analytic results in the literature \cite{Falkowski:2012fb} (dashed curves), for elastic (left panel) and inelastic (right panel) cases (axes are in units of $v_\phi$). For this plot, we have fixed $\lambda=1,~l_w=10 v_\phi,~\gamma=500,$ and $L_p=80/v_\phi$. }
    \label{fig:numberdensities}
\end{figure}

In \Fref{fig:numberdensities}, we plot the number of $\psi$ particles produced per unit area of bubble wall, $n_\psi/A$, for both elastic and inelastic cases as a function of the mass $m_\psi$, fixing $\lambda=1$, calculated using \Eref{numberenergy}. We plot the results obtained with our fit functions as solid curves, and the results obtained from using the analytic formulae from \cite{Falkowski:2012fb} (\Eref{eq:fchielastic} and \Eref{eq:fchiTI}) as dashed curves. 

For elastic collisions, we see that particle production is enhanced at $2 m_\psi\!<\!m_f$ relative to the analytic estimate due to the presence of the peak in the efficiency factor due to oscillations, which is absent in the analytic result, but matches the analytic result at higher masses when the resonance peak becomes kinematically inaccessible. At large masses, the number per unit area scales as $\sim (m/T)^{-2}$, where $T$ is the temperature of the bath (assuming $T\approx v_\phi$). Note that this scaling applies only to the production of scalars (with the decay probability given by \Eref{specific}); for other interaction forms this power dependence can change. 

For inelastic collisions, particle production is also enhanced for $2 m_\psi < m_f$, as the analytic estimate undermines the amplitude of the peak; however, this is now only true for the chosen set of values, and for other parameter choices (in particular a larger value of $\gamma_w/l_w$), it is possible that the analytic results could predict higher particle number densities. The crucial difference is seen at larger masses, where the particle number for the analytic estimate drops far more steeply than for our results. In this large mass regime, our results predict that particle number per unit area scales as $\sim (m/T)^{-2}$ as in the elastic case, in stark contrast to the $\sim (m/T)^{-6}$ scaling predicted by the analytic result.  

It is worth emphasizing that this produced particle number is per unit surface area of bubble walls, and will diffuse into the volume of the bubbles, which should be accounted for in calculations of overall particle number densities. 

\section{Discussions and Summary}
\label{sec:discussion}

In this paper, we performed numerical studies of particle production from realistic bubble collision configurations during first order phase transitions, building on simplified treatments of this effect in either perfectly elastic or totally inelastic configurations in the literature \cite{Watkins:1991zt,Konstandin:2011ds,Falkowski:2012fb}. Our main findings can be summarized as follows:
\begin{itemize}

\item In all realistic scenarios, elastic as well as inelastic, the efficiency factor for particle production $f(\chi)$ (see \Fref{fig:efficiencyfactors}) consists of a power law $f(\chi)\sim \chi^{-2}$ from the collision process between the bubble walls, and a peak around $\chi\sim m_{t,f}^2$ (where $m_t,m_f$ are the scalar field masses in the true and false vacua), corresponding to scalar field oscillations around the true or false minimum after bubble collisions respectively. Simplified treatments in the literature miss the peak for elastic collisions and the power law at high energies for inelastic collisions, but our numerical results indicate that both features exist in realistic collision scenarios. 

\item We characterize the nature of the peak (see Sec.\,\ref{subsec:peak}), clarifying its dependence on physical parameters and phenomena, in particular on the volume of spacetime that the scalar oscillations extend to.  

\item For a simple scalar production scenario (see Sec.\,\ref{particleprod}), we find that the simple analytic estimates in the literature underestimate the production of particles. For particle masses smaller than the mass of the scalar undergoing the phase transition, this is due to the analytic estimates missing or underestimating the peak in the efficiency factor. For masses higher than the scale/temperature of the phase transition, we find that number densities scale as $\sim (m/T)^{-2}$ in all cases. This matches the results in the literature for elastic collisions, and is far stronger than the $\sim (m/T)^{-6}$ scaling found in \cite{Falkowski:2012fb} for totally inelastic collisions. Notably, this gives a significantly larger particle abundance than the familiar $e^{-m/T}$ Boltzmann suppressed abundance from thermal processes for $m/T>\mathcal{O}(10)$. Our results therefore confirm that collisions of boosted bubble walls can be an efficient source of heavy particles. Moreover, our results show that this is true for runaway bubble collisions in general, independent of whether they are elastic or inelastic. 

\item We provide easy to use analytic formulae that fit our numerical results: \Eref{eq:elasticfit} for elastic collisions, \Eref{eq:inelasticfit} for inelastic collisions, together with \Eref{eq:fchielastic}. These are valid for energies between the inverse bubble radius at collision $R_*^{-1}$ and the inverse boosted bubble wall thickness $\gamma_w/l_w$, and can be used to estimate particle production in generic bubble collision setups.

\end{itemize}

In this paper, we have focused on improving the results for the efficiency factor for various bubble collision configurations through numerical studies. More detailed discussions of the underlying physics, including understanding the relative contributions from various stages of the phase transition, the physical aspects of the collision process, and possible importance of resonant effects are discussed in a companion paper \cite{Shakya:2023kjf}. These results have since been applied to various BSM phenomena, such as dark matter \cite{Giudice:2024tcp} and leptogenesis \cite{Cataldi:2024pgt}. Copious particle production from bubble collisions can also modify the gravitational wave signals from such phase transitions \cite{Inomata:2024rkt}. 

Let us briefly discuss some limitations of our results. Our results were derived for a specific parameterization of the scalar potential \Eref{eq:toypotential}. It is possible that a potential that is very different could produce results that can deviate from our findings; however, the qualitative form of our results (discussed above) is robust, and should continue to hold. Our numerical studies were also carried out for a single collision of two planar walls over a spacetime region significantly smaller than the size of a typical bubble at collision. More realistic scenarios involving multiple collisions (as can occur for the elastic case), or multiple bubbles of different sizes, could introduce additional physical features that might be relevant, but are beyond the scope of this paper. Furthermore, we have ignored the backreaction from particle production on the dynamics of the scalar field, but this could be relevant if particle production is a very strong effect. All of these represent interesting directions for further careful study.

\acknowledgments
We are especially grateful to Thomas Konstandin and Geraldine Servant for several detailed discussions on various aspects of this work. We also acknowledge illuminating conversations with Gian Giudice, Ryusuke Jinno, Hyungjin Kim, Hyun Min Lee, Alex Pomarol, and Jorinde van de Vis. BS is supported by the Deutsche Forschungsgemeinschaft under Germany’s Excellence Strategy - EXC 2121 Quantum Universe - 390833306. HM is supported by the Deutsche Forschungsgemeinschaft (DFG, German Research Foundation) under grant 396021762 - TRR 257.

\appendix

\section{Spurious Contributions and Imaginary Components}
\label{app:imaginary}

\begin{figure}[h]
    \centering   \includegraphics[width=1\linewidth]{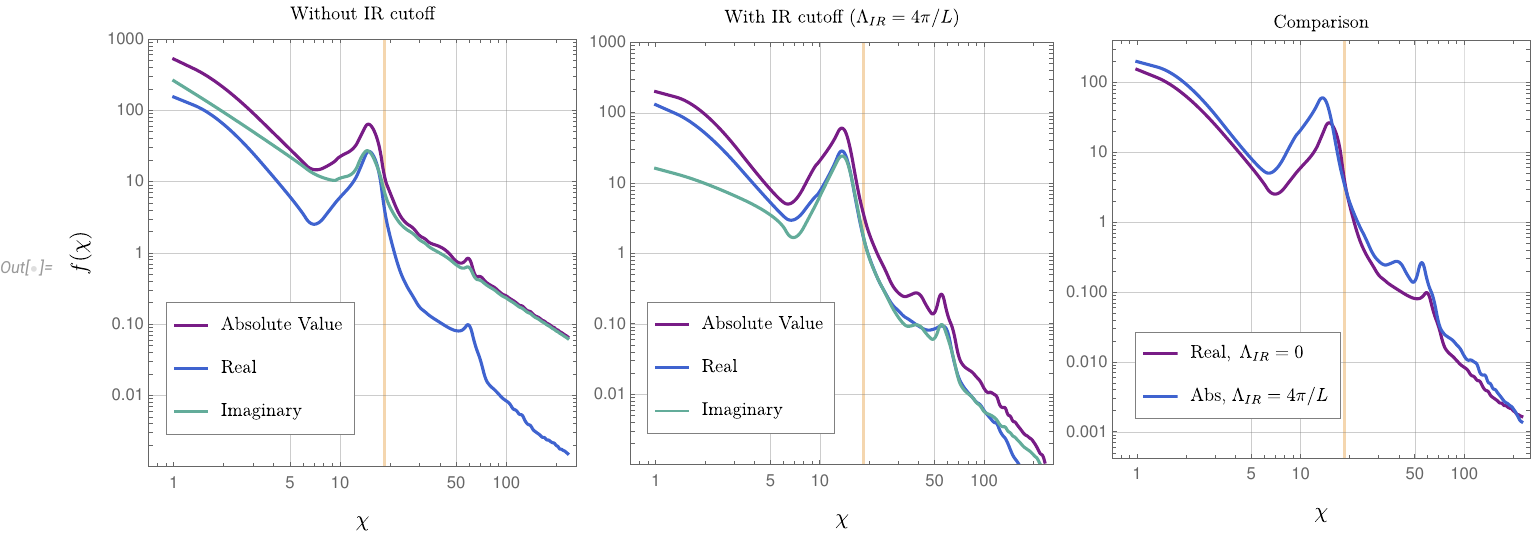}
    \caption{Different components of the efficiency factor for $\epsilon=0.06$. We compute separately the efficiency factor from the real and imaginary parts of the Fourier transform, as well as its absolute value. Left: straightforward numerical computation without imposing an IR cutoff. Center: Imposing a cutoff for contributions from modes with $\omega,k<4\pi/L$. Right: comparison between the real contribution without IR cutoff and the absolute value with the IR cutoff imposed, showing excellent agreement.}
    \label{fig:fullresults}
\end{figure}

As mentioned in the text, the finite size of our numerical studies (size $L$) can incur spurious effects close to the lower frequency and momentum cutoffs (i.e.\,low $\omega, k$ values $\sim 2\pi/L$). To avoid such effects, one should take an infrared (IR) cutoff close to $\omega, k\sim 2\pi/L$. However, this IR cutoff is not straightforward to incorporate in our numerical studies, since the efficiency factor is calculated in terms of the variables $\xi=\omega^2+k^2$ and  $\chi=\omega^2-k^2$ in order to facilitate direct comparisons with previous results in the literature. However, these spurious contributions are correlated with imaginary components of the Fourier transform. This is illustrated for a sample case in Fig.\,\ref{fig:fullresults}.

The left panel shows the efficiency factor as calculated using the formalism discussed in the main text, without taking any IR cutoff; we plot the absolute value of the function, as well as its real and imaginary components, which shows that the function is clearly dominated by its imaginary component. In the central panel, we impose the IR cutoff by manually removing the contributions corresponding to $\omega,k<4\pi/L$ before performing the change of variables and computing the efficiency factor. This shows that while real component remains more-or-less unchanged compared to the left panel, the imaginary component gets significantly suppressed and no longer provides the dominant contribution.  In the third panel, we overlay the real part of the efficiency factor calculated without the IR cutoff with the absolute value of the corresponding computation with the IR cutoff imposed; these are found to agree within a factor of $2$. We have checked several other cases (corresponding to different values of $\epsilon$), and found that these correlations persists for all checked cases. Based on these obervations, in our calculations we therefore take only the real part of the Fourier transform, which is computationally easier to implement, instead of taking an IR cutoff for the $\omega,k$ values.

\section{Comparison between Numerical Results and Fit Formulae}
\label{app:fits}

\begin{figure}
    \centering
    \includegraphics[width=0.6\textwidth]{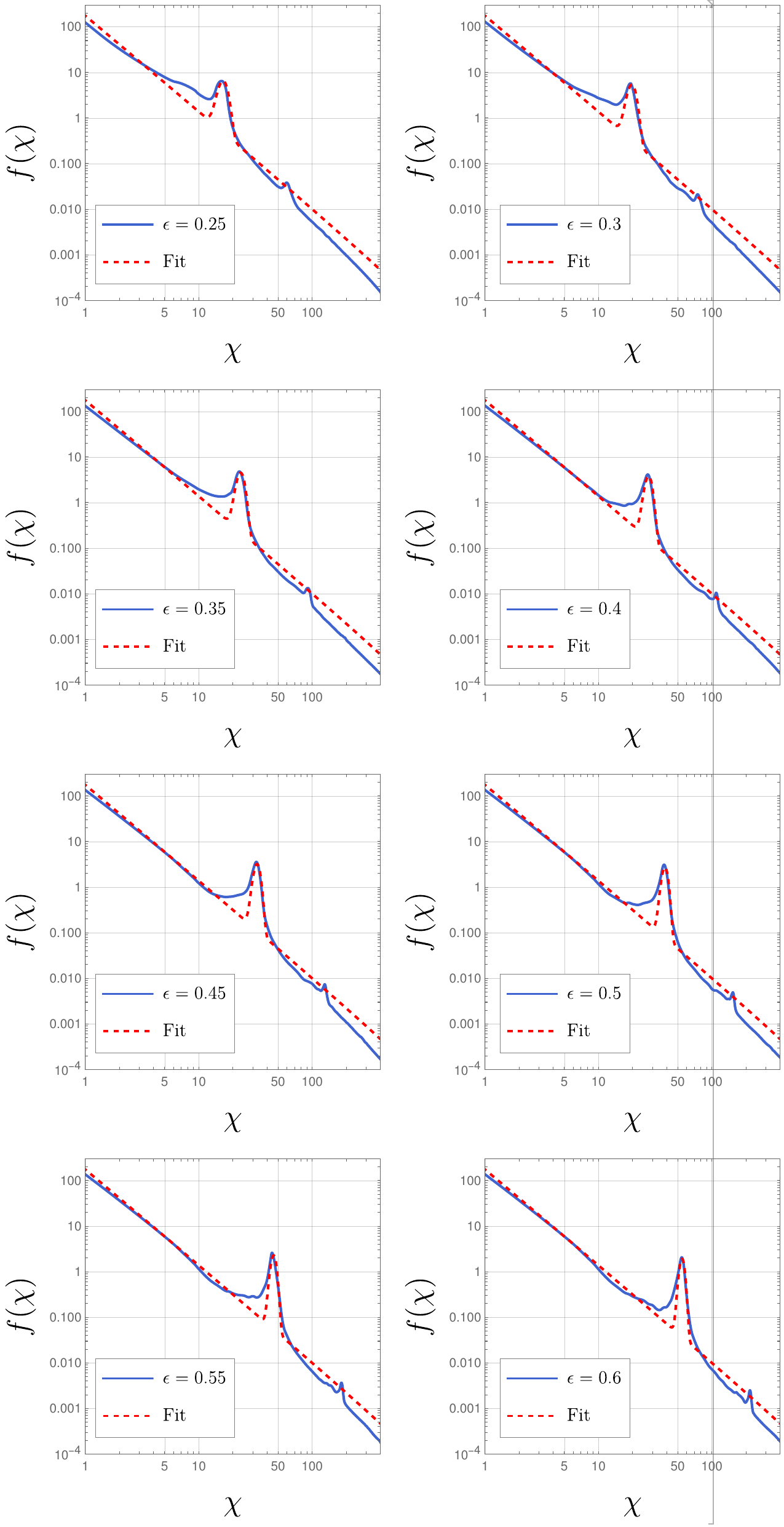}
    \caption{Comparison between our numerical results and the analytic fit formula \Eref{eq:elasticfit} for elastic collisions. }
    \label{fig:app-elas-fits}
\end{figure}

\begin{figure}
    \centering
    \includegraphics[width=0.6\textwidth]{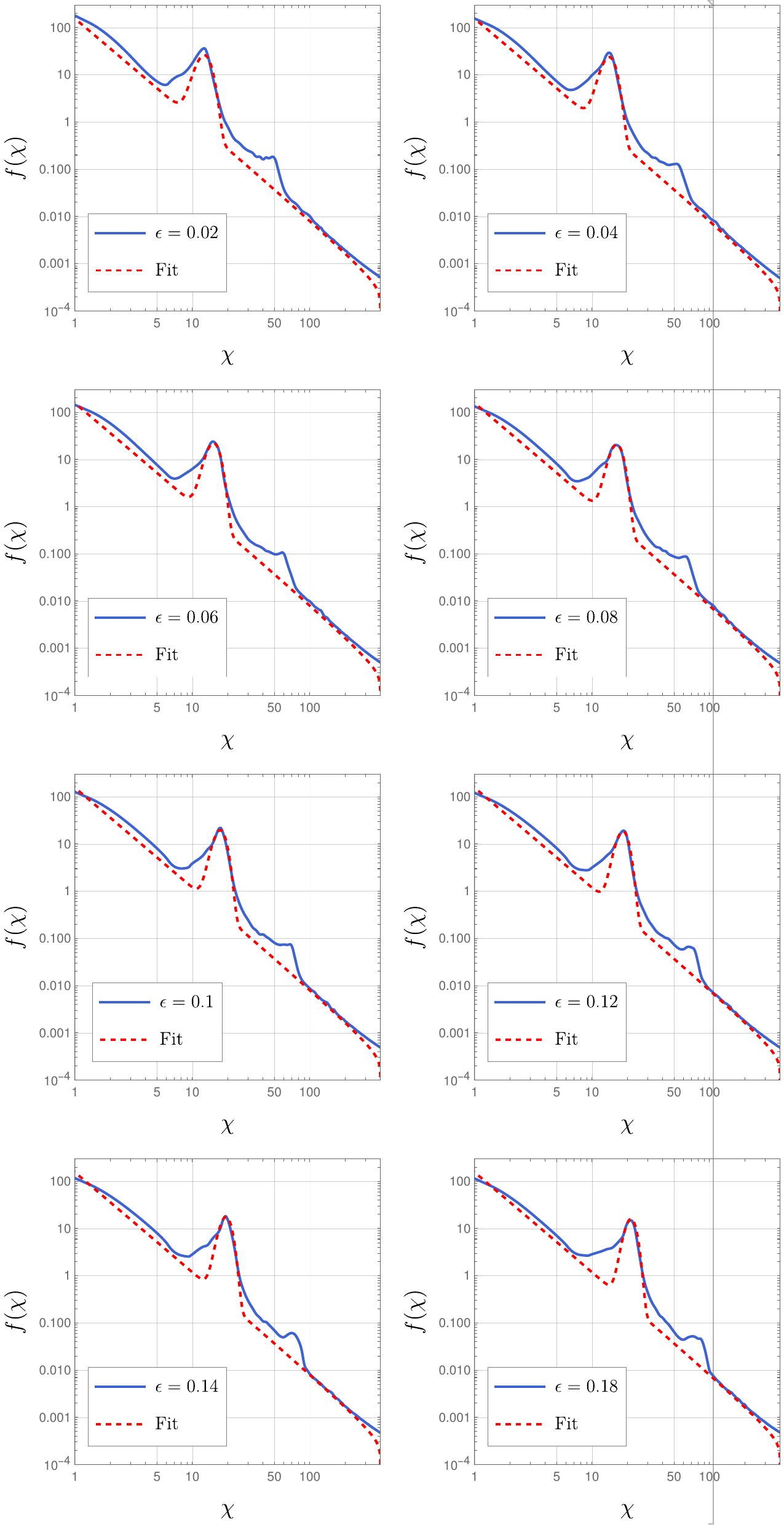}
    \caption{Comparison between our numerical results and the analytic fit formula \Eref{eq:inelasticfit} for inelastic collisions.}
    \label{fig:app-inelas-fits}
\end{figure}

Here we plot the comparisons between our numerical results and our analytic fit formulae \Eref{eq:elasticfit}, \Eref{eq:inelasticfit} for different potential shapes (i.e.\,different values of $\epsilon$) for elastic (\Fref{fig:app-elas-fits}) and inelastic (\Fref{fig:app-inelas-fits}) collisions. For these plots, we have fixed $\gamma_{w}=200$, $l_{w}=10/v_{\phi}$, and $L_p=40/v_{\phi}$.

\bibliography{references}

%apsrev4-2.bst 2019-01-14 (MD) hand-edited version of apsrev4-1.bst
%Control: key (0)
%Control: author (8) initials jnrlst
%Control: editor formatted (1) identically to author
%Control: production of article title (0) allowed
%Control: page (0) single
%Control: year (1) truncated
%Control: production of eprint (0) enabled
\begin{thebibliography}{44}%
\makeatletter
\providecommand \@ifxundefined [1]{%
 \@ifx{#1\undefined}
}%
\providecommand \@ifnum [1]{%
 \ifnum #1\expandafter \@firstoftwo
 \else \expandafter \@secondoftwo
 \fi
}%
\providecommand \@ifx [1]{%
 \ifx #1\expandafter \@firstoftwo
 \else \expandafter \@secondoftwo
 \fi
}%
\providecommand \natexlab [1]{#1}%
\providecommand \enquote  [1]{``#1''}%
\providecommand \bibnamefont  [1]{#1}%
\providecommand \bibfnamefont [1]{#1}%
\providecommand \citenamefont [1]{#1}%
\providecommand \href@noop [0]{\@secondoftwo}%
\providecommand \href [0]{\begingroup \@sanitize@url \@href}%
\providecommand \@href[1]{\@@startlink{#1}\@@href}%
\providecommand \@@href[1]{\endgroup#1\@@endlink}%
\providecommand \@sanitize@url [0]{\catcode `\\12\catcode `\$12\catcode
  `\&12\catcode `\#12\catcode `\^12\catcode `\_12\catcode `\%12\relax}%
\providecommand \@@startlink[1]{}%
\providecommand \@@endlink[0]{}%
\providecommand \url  [0]{\begingroup\@sanitize@url \@url }%
\providecommand \@url [1]{\endgroup\@href {#1}{\urlprefix }}%
\providecommand \urlprefix  [0]{URL }%
\providecommand \Eprint [0]{\href }%
\providecommand \doibase [0]{https://doi.org/}%
\providecommand \selectlanguage [0]{\@gobble}%
\providecommand \bibinfo  [0]{\@secondoftwo}%
\providecommand \bibfield  [0]{\@secondoftwo}%
\providecommand \translation [1]{[#1]}%
\providecommand \BibitemOpen [0]{}%
\providecommand \bibitemStop [0]{}%
\providecommand \bibitemNoStop [0]{.\EOS\space}%
\providecommand \EOS [0]{\spacefactor3000\relax}%
\providecommand \BibitemShut  [1]{\csname bibitem#1\endcsname}%
\let\auto@bib@innerbib\@empty
%</preamble>
\bibitem [{\citenamefont {Hogan}(1983)}]{Hogan:1983ixn}%
  \BibitemOpen
  \bibfield  {author} {\bibinfo {author} {\bibfnamefont {C.~J.}\ \bibnamefont
  {Hogan}},\ }\bibfield  {title} {\bibinfo {title} {{NUCLEATION OF COSMOLOGICAL
  PHASE TRANSITIONS}},\ }\href {https://doi.org/10.1016/0370-2693(83)90553-1}
  {\bibfield  {journal} {\bibinfo  {journal} {Phys. Lett. B}\ }\textbf
  {\bibinfo {volume} {133}},\ \bibinfo {pages} {172} (\bibinfo {year}
  {1983})}\BibitemShut {NoStop}%
\bibitem [{\citenamefont {Witten}(1984)}]{Witten:1984rs}%
  \BibitemOpen
  \bibfield  {author} {\bibinfo {author} {\bibfnamefont {E.}~\bibnamefont
  {Witten}},\ }\bibfield  {title} {\bibinfo {title} {{Cosmic Separation of
  Phases}},\ }\href {https://doi.org/10.1103/PhysRevD.30.272} {\bibfield
  {journal} {\bibinfo  {journal} {Phys. Rev. D}\ }\textbf {\bibinfo {volume}
  {30}},\ \bibinfo {pages} {272} (\bibinfo {year} {1984})}\BibitemShut
  {NoStop}%
\bibitem [{\citenamefont {Hogan}(1986)}]{Hogan:1986qda}%
  \BibitemOpen
  \bibfield  {author} {\bibinfo {author} {\bibfnamefont {C.~J.}\ \bibnamefont
  {Hogan}},\ }\bibfield  {title} {\bibinfo {title} {{Gravitational radiation
  from cosmological phase transitions}},\ }\href@noop {} {\bibfield  {journal}
  {\bibinfo  {journal} {Mon. Not. Roy. Astron. Soc.}\ }\textbf {\bibinfo
  {volume} {218}},\ \bibinfo {pages} {629} (\bibinfo {year}
  {1986})}\BibitemShut {NoStop}%
\bibitem [{\citenamefont {Kosowsky}\ \emph
  {et~al.}(1992{\natexlab{a}})\citenamefont {Kosowsky}, \citenamefont
  {Turner},\ and\ \citenamefont {Watkins}}]{Kosowsky:1991ua}%
  \BibitemOpen
  \bibfield  {author} {\bibinfo {author} {\bibfnamefont {A.}~\bibnamefont
  {Kosowsky}}, \bibinfo {author} {\bibfnamefont {M.~S.}\ \bibnamefont
  {Turner}},\ and\ \bibinfo {author} {\bibfnamefont {R.}~\bibnamefont
  {Watkins}},\ }\bibfield  {title} {\bibinfo {title} {{Gravitational radiation
  from colliding vacuum bubbles}},\ }\href
  {https://doi.org/10.1103/PhysRevD.45.4514} {\bibfield  {journal} {\bibinfo
  {journal} {Phys. Rev. D}\ }\textbf {\bibinfo {volume} {45}},\ \bibinfo
  {pages} {4514} (\bibinfo {year} {1992}{\natexlab{a}})}\BibitemShut {NoStop}%
\bibitem [{\citenamefont {Kosowsky}\ \emph
  {et~al.}(1992{\natexlab{b}})\citenamefont {Kosowsky}, \citenamefont
  {Turner},\ and\ \citenamefont {Watkins}}]{Kosowsky:1992rz}%
  \BibitemOpen
  \bibfield  {author} {\bibinfo {author} {\bibfnamefont {A.}~\bibnamefont
  {Kosowsky}}, \bibinfo {author} {\bibfnamefont {M.~S.}\ \bibnamefont
  {Turner}},\ and\ \bibinfo {author} {\bibfnamefont {R.}~\bibnamefont
  {Watkins}},\ }\bibfield  {title} {\bibinfo {title} {{Gravitational waves from
  first order cosmological phase transitions}},\ }\href
  {https://doi.org/10.1103/PhysRevLett.69.2026} {\bibfield  {journal} {\bibinfo
   {journal} {Phys. Rev. Lett.}\ }\textbf {\bibinfo {volume} {69}},\ \bibinfo
  {pages} {2026} (\bibinfo {year} {1992}{\natexlab{b}})}\BibitemShut {NoStop}%
\bibitem [{\citenamefont {Kosowsky}\ and\ \citenamefont
  {Turner}(1993)}]{Kosowsky:1992vn}%
  \BibitemOpen
  \bibfield  {author} {\bibinfo {author} {\bibfnamefont {A.}~\bibnamefont
  {Kosowsky}}\ and\ \bibinfo {author} {\bibfnamefont {M.~S.}\ \bibnamefont
  {Turner}},\ }\bibfield  {title} {\bibinfo {title} {{Gravitational radiation
  from colliding vacuum bubbles: envelope approximation to many bubble
  collisions}},\ }\href {https://doi.org/10.1103/PhysRevD.47.4372} {\bibfield
  {journal} {\bibinfo  {journal} {Phys. Rev. D}\ }\textbf {\bibinfo {volume}
  {47}},\ \bibinfo {pages} {4372} (\bibinfo {year} {1993})},\ \Eprint
  {https://arxiv.org/abs/astro-ph/9211004} {arXiv:astro-ph/9211004}
  \BibitemShut {NoStop}%
\bibitem [{\citenamefont {Kamionkowski}\ \emph {et~al.}(1994)\citenamefont
  {Kamionkowski}, \citenamefont {Kosowsky},\ and\ \citenamefont
  {Turner}}]{Kamionkowski:1993fg}%
  \BibitemOpen
  \bibfield  {author} {\bibinfo {author} {\bibfnamefont {M.}~\bibnamefont
  {Kamionkowski}}, \bibinfo {author} {\bibfnamefont {A.}~\bibnamefont
  {Kosowsky}},\ and\ \bibinfo {author} {\bibfnamefont {M.~S.}\ \bibnamefont
  {Turner}},\ }\bibfield  {title} {\bibinfo {title} {{Gravitational radiation
  from first order phase transitions}},\ }\href
  {https://doi.org/10.1103/PhysRevD.49.2837} {\bibfield  {journal} {\bibinfo
  {journal} {Phys. Rev. D}\ }\textbf {\bibinfo {volume} {49}},\ \bibinfo
  {pages} {2837} (\bibinfo {year} {1994})},\ \Eprint
  {https://arxiv.org/abs/astro-ph/9310044} {arXiv:astro-ph/9310044}
  \BibitemShut {NoStop}%
\bibitem [{\citenamefont {Schwaller}(2015)}]{Schwaller:2015tja}%
  \BibitemOpen
  \bibfield  {author} {\bibinfo {author} {\bibfnamefont {P.}~\bibnamefont
  {Schwaller}},\ }\bibfield  {title} {\bibinfo {title} {{Gravitational Waves
  from a Dark Phase Transition}},\ }\href
  {https://doi.org/10.1103/PhysRevLett.115.181101} {\bibfield  {journal}
  {\bibinfo  {journal} {Phys. Rev. Lett.}\ }\textbf {\bibinfo {volume} {115}},\
  \bibinfo {pages} {181101} (\bibinfo {year} {2015})},\ \Eprint
  {https://arxiv.org/abs/1504.07263} {arXiv:1504.07263 [hep-ph]} \BibitemShut
  {NoStop}%
\bibitem [{\citenamefont {Jaeckel}\ \emph {et~al.}(2016)\citenamefont
  {Jaeckel}, \citenamefont {Khoze},\ and\ \citenamefont
  {Spannowsky}}]{Jaeckel:2016jlh}%
  \BibitemOpen
  \bibfield  {author} {\bibinfo {author} {\bibfnamefont {J.}~\bibnamefont
  {Jaeckel}}, \bibinfo {author} {\bibfnamefont {V.~V.}\ \bibnamefont {Khoze}},\
  and\ \bibinfo {author} {\bibfnamefont {M.}~\bibnamefont {Spannowsky}},\
  }\bibfield  {title} {\bibinfo {title} {{Hearing the signal of dark sectors
  with gravitational wave detectors}},\ }\href
  {https://doi.org/10.1103/PhysRevD.94.103519} {\bibfield  {journal} {\bibinfo
  {journal} {Phys. Rev. D}\ }\textbf {\bibinfo {volume} {94}},\ \bibinfo
  {pages} {103519} (\bibinfo {year} {2016})},\ \Eprint
  {https://arxiv.org/abs/1602.03901} {arXiv:1602.03901 [hep-ph]} \BibitemShut
  {NoStop}%
\bibitem [{\citenamefont {Dev}\ and\ \citenamefont
  {Mazumdar}(2016)}]{Dev:2016feu}%
  \BibitemOpen
  \bibfield  {author} {\bibinfo {author} {\bibfnamefont {P.~S.~B.}\
  \bibnamefont {Dev}}\ and\ \bibinfo {author} {\bibfnamefont {A.}~\bibnamefont
  {Mazumdar}},\ }\bibfield  {title} {\bibinfo {title} {{Probing the Scale of
  New Physics by Advanced LIGO/VIRGO}},\ }\href
  {https://doi.org/10.1103/PhysRevD.93.104001} {\bibfield  {journal} {\bibinfo
  {journal} {Phys. Rev. D}\ }\textbf {\bibinfo {volume} {93}},\ \bibinfo
  {pages} {104001} (\bibinfo {year} {2016})},\ \Eprint
  {https://arxiv.org/abs/1602.04203} {arXiv:1602.04203 [hep-ph]} \BibitemShut
  {NoStop}%
\bibitem [{\citenamefont {Baldes}(2017)}]{Baldes:2017rcu}%
  \BibitemOpen
  \bibfield  {author} {\bibinfo {author} {\bibfnamefont {I.}~\bibnamefont
  {Baldes}},\ }\bibfield  {title} {\bibinfo {title} {{Gravitational waves from
  the asymmetric-dark-matter generating phase transition}},\ }\href
  {https://doi.org/10.1088/1475-7516/2017/05/028} {\bibfield  {journal}
  {\bibinfo  {journal} {JCAP}\ }\textbf {\bibinfo {volume} {05}},\ \bibinfo
  {pages} {028}},\ \Eprint {https://arxiv.org/abs/1702.02117} {arXiv:1702.02117
  [hep-ph]} \BibitemShut {NoStop}%
\bibitem [{\citenamefont {Tsumura}\ \emph {et~al.}(2017)\citenamefont
  {Tsumura}, \citenamefont {Yamada},\ and\ \citenamefont
  {Yamaguchi}}]{Tsumura:2017knk}%
  \BibitemOpen
  \bibfield  {author} {\bibinfo {author} {\bibfnamefont {K.}~\bibnamefont
  {Tsumura}}, \bibinfo {author} {\bibfnamefont {M.}~\bibnamefont {Yamada}},\
  and\ \bibinfo {author} {\bibfnamefont {Y.}~\bibnamefont {Yamaguchi}},\
  }\bibfield  {title} {\bibinfo {title} {{Gravitational wave from dark sector
  with dark pion}},\ }\href {https://doi.org/10.1088/1475-7516/2017/07/044}
  {\bibfield  {journal} {\bibinfo  {journal} {JCAP}\ }\textbf {\bibinfo
  {volume} {07}},\ \bibinfo {pages} {044}},\ \Eprint
  {https://arxiv.org/abs/1704.00219} {arXiv:1704.00219 [hep-ph]} \BibitemShut
  {NoStop}%
\bibitem [{\citenamefont {Okada}\ and\ \citenamefont
  {Seto}(2018)}]{Okada:2018xdh}%
  \BibitemOpen
  \bibfield  {author} {\bibinfo {author} {\bibfnamefont {N.}~\bibnamefont
  {Okada}}\ and\ \bibinfo {author} {\bibfnamefont {O.}~\bibnamefont {Seto}},\
  }\bibfield  {title} {\bibinfo {title} {{Probing the seesaw scale with
  gravitational waves}},\ }\href {https://doi.org/10.1103/PhysRevD.98.063532}
  {\bibfield  {journal} {\bibinfo  {journal} {Phys. Rev. D}\ }\textbf {\bibinfo
  {volume} {98}},\ \bibinfo {pages} {063532} (\bibinfo {year} {2018})},\
  \Eprint {https://arxiv.org/abs/1807.00336} {arXiv:1807.00336 [hep-ph]}
  \BibitemShut {NoStop}%
\bibitem [{\citenamefont {Croon}\ \emph {et~al.}(2018)\citenamefont {Croon},
  \citenamefont {Sanz},\ and\ \citenamefont {White}}]{Croon:2018erz}%
  \BibitemOpen
  \bibfield  {author} {\bibinfo {author} {\bibfnamefont {D.}~\bibnamefont
  {Croon}}, \bibinfo {author} {\bibfnamefont {V.}~\bibnamefont {Sanz}},\ and\
  \bibinfo {author} {\bibfnamefont {G.}~\bibnamefont {White}},\ }\bibfield
  {title} {\bibinfo {title} {{Model Discrimination in Gravitational Wave
  spectra from Dark Phase Transitions}},\ }\href
  {https://doi.org/10.1007/JHEP08(2018)203} {\bibfield  {journal} {\bibinfo
  {journal} {JHEP}\ }\textbf {\bibinfo {volume} {08}},\ \bibinfo {pages}
  {203}},\ \Eprint {https://arxiv.org/abs/1806.02332} {arXiv:1806.02332
  [hep-ph]} \BibitemShut {NoStop}%
\bibitem [{\citenamefont {Baldes}\ and\ \citenamefont
  {Garcia-Cely}(2019)}]{Baldes:2018emh}%
  \BibitemOpen
  \bibfield  {author} {\bibinfo {author} {\bibfnamefont {I.}~\bibnamefont
  {Baldes}}\ and\ \bibinfo {author} {\bibfnamefont {C.}~\bibnamefont
  {Garcia-Cely}},\ }\bibfield  {title} {\bibinfo {title} {{Strong gravitational
  radiation from a simple dark matter model}},\ }\href
  {https://doi.org/10.1007/JHEP05(2019)190} {\bibfield  {journal} {\bibinfo
  {journal} {JHEP}\ }\textbf {\bibinfo {volume} {05}},\ \bibinfo {pages}
  {190}},\ \Eprint {https://arxiv.org/abs/1809.01198} {arXiv:1809.01198
  [hep-ph]} \BibitemShut {NoStop}%
\bibitem [{\citenamefont {Prokopec}\ \emph {et~al.}(2019)\citenamefont
  {Prokopec}, \citenamefont {Rezacek},\ and\ \citenamefont
  {\'Swie\.zewska}}]{Prokopec:2018tnq}%
  \BibitemOpen
  \bibfield  {author} {\bibinfo {author} {\bibfnamefont {T.}~\bibnamefont
  {Prokopec}}, \bibinfo {author} {\bibfnamefont {J.}~\bibnamefont {Rezacek}},\
  and\ \bibinfo {author} {\bibfnamefont {B.}~\bibnamefont {\'Swie\.zewska}},\
  }\bibfield  {title} {\bibinfo {title} {{Gravitational waves from conformal
  symmetry breaking}},\ }\href {https://doi.org/10.1088/1475-7516/2019/02/009}
  {\bibfield  {journal} {\bibinfo  {journal} {JCAP}\ }\textbf {\bibinfo
  {volume} {02}},\ \bibinfo {pages} {009}},\ \Eprint
  {https://arxiv.org/abs/1809.11129} {arXiv:1809.11129 [hep-ph]} \BibitemShut
  {NoStop}%
\bibitem [{\citenamefont {Bai}\ \emph {et~al.}(2019)\citenamefont {Bai},
  \citenamefont {Long},\ and\ \citenamefont {Lu}}]{Bai:2018dxf}%
  \BibitemOpen
  \bibfield  {author} {\bibinfo {author} {\bibfnamefont {Y.}~\bibnamefont
  {Bai}}, \bibinfo {author} {\bibfnamefont {A.~J.}\ \bibnamefont {Long}},\ and\
  \bibinfo {author} {\bibfnamefont {S.}~\bibnamefont {Lu}},\ }\bibfield
  {title} {\bibinfo {title} {{Dark Quark Nuggets}},\ }\href
  {https://doi.org/10.1103/PhysRevD.99.055047} {\bibfield  {journal} {\bibinfo
  {journal} {Phys. Rev. D}\ }\textbf {\bibinfo {volume} {99}},\ \bibinfo
  {pages} {055047} (\bibinfo {year} {2019})},\ \Eprint
  {https://arxiv.org/abs/1810.04360} {arXiv:1810.04360 [hep-ph]} \BibitemShut
  {NoStop}%
\bibitem [{\citenamefont {Breitbach}\ \emph {et~al.}(2019)\citenamefont
  {Breitbach}, \citenamefont {Kopp}, \citenamefont {Madge}, \citenamefont
  {Opferkuch},\ and\ \citenamefont {Schwaller}}]{Breitbach:2018ddu}%
  \BibitemOpen
  \bibfield  {author} {\bibinfo {author} {\bibfnamefont {M.}~\bibnamefont
  {Breitbach}}, \bibinfo {author} {\bibfnamefont {J.}~\bibnamefont {Kopp}},
  \bibinfo {author} {\bibfnamefont {E.}~\bibnamefont {Madge}}, \bibinfo
  {author} {\bibfnamefont {T.}~\bibnamefont {Opferkuch}},\ and\ \bibinfo
  {author} {\bibfnamefont {P.}~\bibnamefont {Schwaller}},\ }\bibfield  {title}
  {\bibinfo {title} {{Dark, Cold, and Noisy: Constraining Secluded Hidden
  Sectors with Gravitational Waves}},\ }\href
  {https://doi.org/10.1088/1475-7516/2019/07/007} {\bibfield  {journal}
  {\bibinfo  {journal} {JCAP}\ }\textbf {\bibinfo {volume} {07}},\ \bibinfo
  {pages} {007}},\ \Eprint {https://arxiv.org/abs/1811.11175} {arXiv:1811.11175
  [hep-ph]} \BibitemShut {NoStop}%
\bibitem [{\citenamefont {Fairbairn}\ \emph {et~al.}(2019)\citenamefont
  {Fairbairn}, \citenamefont {Hardy},\ and\ \citenamefont
  {Wickens}}]{Fairbairn:2019xog}%
  \BibitemOpen
  \bibfield  {author} {\bibinfo {author} {\bibfnamefont {M.}~\bibnamefont
  {Fairbairn}}, \bibinfo {author} {\bibfnamefont {E.}~\bibnamefont {Hardy}},\
  and\ \bibinfo {author} {\bibfnamefont {A.}~\bibnamefont {Wickens}},\
  }\bibfield  {title} {\bibinfo {title} {{Hearing without seeing: gravitational
  waves from hot and cold hidden sectors}},\ }\href
  {https://doi.org/10.1007/JHEP07(2019)044} {\bibfield  {journal} {\bibinfo
  {journal} {JHEP}\ }\textbf {\bibinfo {volume} {07}},\ \bibinfo {pages}
  {044}},\ \Eprint {https://arxiv.org/abs/1901.11038} {arXiv:1901.11038
  [hep-ph]} \BibitemShut {NoStop}%
\bibitem [{\citenamefont {Helmboldt}\ \emph {et~al.}(2019)\citenamefont
  {Helmboldt}, \citenamefont {Kubo},\ and\ \citenamefont {van~der
  Woude}}]{Helmboldt:2019pan}%
  \BibitemOpen
  \bibfield  {author} {\bibinfo {author} {\bibfnamefont {A.~J.}\ \bibnamefont
  {Helmboldt}}, \bibinfo {author} {\bibfnamefont {J.}~\bibnamefont {Kubo}},\
  and\ \bibinfo {author} {\bibfnamefont {S.}~\bibnamefont {van~der Woude}},\
  }\bibfield  {title} {\bibinfo {title} {{Observational prospects for
  gravitational waves from hidden or dark chiral phase transitions}},\ }\href
  {https://doi.org/10.1103/PhysRevD.100.055025} {\bibfield  {journal} {\bibinfo
   {journal} {Phys. Rev. D}\ }\textbf {\bibinfo {volume} {100}},\ \bibinfo
  {pages} {055025} (\bibinfo {year} {2019})},\ \Eprint
  {https://arxiv.org/abs/1904.07891} {arXiv:1904.07891 [hep-ph]} \BibitemShut
  {NoStop}%
\bibitem [{\citenamefont {Ertas}\ \emph {et~al.}(2022)\citenamefont {Ertas},
  \citenamefont {Kahlhoefer},\ and\ \citenamefont {Tasillo}}]{Ertas:2021xeh}%
  \BibitemOpen
  \bibfield  {author} {\bibinfo {author} {\bibfnamefont {F.}~\bibnamefont
  {Ertas}}, \bibinfo {author} {\bibfnamefont {F.}~\bibnamefont {Kahlhoefer}},\
  and\ \bibinfo {author} {\bibfnamefont {C.}~\bibnamefont {Tasillo}},\
  }\bibfield  {title} {\bibinfo {title} {{Turn up the volume: listening to
  phase transitions in hot dark sectors}},\ }\href
  {https://doi.org/10.1088/1475-7516/2022/02/014} {\bibfield  {journal}
  {\bibinfo  {journal} {JCAP}\ }\textbf {\bibinfo {volume} {02}}\bibfield
  {number} {\bibinfo  {number} { (02)},\ \bibinfo {pages} {014}},\ }\Eprint
  {https://arxiv.org/abs/2109.06208} {arXiv:2109.06208 [astro-ph.CO]}
  \BibitemShut {NoStop}%
\bibitem [{\citenamefont {Jinno}\ \emph {et~al.}(2022)\citenamefont {Jinno},
  \citenamefont {Shakya},\ and\ \citenamefont {van~de Vis}}]{Jinno:2022fom}%
  \BibitemOpen
  \bibfield  {author} {\bibinfo {author} {\bibfnamefont {R.}~\bibnamefont
  {Jinno}}, \bibinfo {author} {\bibfnamefont {B.}~\bibnamefont {Shakya}},\ and\
  \bibinfo {author} {\bibfnamefont {J.}~\bibnamefont {van~de Vis}},\ }\bibfield
   {title} {\bibinfo {title} {{Gravitational Waves from Feebly Interacting
  Particles in a First Order Phase Transition}},\ }\href@noop {} {\  (\bibinfo
  {year} {2022})},\ \Eprint {https://arxiv.org/abs/2211.06405}
  {arXiv:2211.06405 [gr-qc]} \BibitemShut {NoStop}%
\bibitem [{\citenamefont {Grojean}\ and\ \citenamefont
  {Servant}(2007)}]{Grojean:2006bp}%
  \BibitemOpen
  \bibfield  {author} {\bibinfo {author} {\bibfnamefont {C.}~\bibnamefont
  {Grojean}}\ and\ \bibinfo {author} {\bibfnamefont {G.}~\bibnamefont
  {Servant}},\ }\bibfield  {title} {\bibinfo {title} {{Gravitational Waves from
  Phase Transitions at the Electroweak Scale and Beyond}},\ }\href
  {https://doi.org/10.1103/PhysRevD.75.043507} {\bibfield  {journal} {\bibinfo
  {journal} {Phys. Rev.}\ }\textbf {\bibinfo {volume} {D75}},\ \bibinfo {pages}
  {043507} (\bibinfo {year} {2007})},\ \Eprint
  {https://arxiv.org/abs/hep-ph/0607107} {arXiv:hep-ph/0607107 [hep-ph]}
  \BibitemShut {NoStop}%
%%CITATION = HEP-PH/0607107;%%
\bibitem [{\citenamefont {Caprini}\ \emph {et~al.}(2016)\citenamefont {Caprini}
  \emph {et~al.}}]{Caprini:2015zlo}%
  \BibitemOpen
  \bibfield  {author} {\bibinfo {author} {\bibfnamefont {C.}~\bibnamefont
  {Caprini}} \emph {et~al.},\ }\bibfield  {title} {\bibinfo {title} {{Science
  with the space-based interferometer eLISA. II: Gravitational waves from
  cosmological phase transitions}},\ }\href
  {https://doi.org/10.1088/1475-7516/2016/04/001} {\bibfield  {journal}
  {\bibinfo  {journal} {JCAP}\ }\textbf {\bibinfo {volume} {1604}},\ \bibinfo
  {pages} {001}},\ \Eprint {https://arxiv.org/abs/1512.06239} {arXiv:1512.06239
  [astro-ph.CO]} \BibitemShut {NoStop}%
%%CITATION = ARXIV:1512.06239;%%
\bibitem [{\citenamefont {Caprini}\ and\ \citenamefont
  {Figueroa}(2018)}]{Caprini:2018mtu}%
  \BibitemOpen
  \bibfield  {author} {\bibinfo {author} {\bibfnamefont {C.}~\bibnamefont
  {Caprini}}\ and\ \bibinfo {author} {\bibfnamefont {D.~G.}\ \bibnamefont
  {Figueroa}},\ }\bibfield  {title} {\bibinfo {title} {{Cosmological
  Backgrounds of Gravitational Waves}},\ }\href
  {https://doi.org/10.1088/1361-6382/aac608} {\bibfield  {journal} {\bibinfo
  {journal} {Class. Quant. Grav.}\ }\textbf {\bibinfo {volume} {35}},\ \bibinfo
  {pages} {163001} (\bibinfo {year} {2018})},\ \Eprint
  {https://arxiv.org/abs/1801.04268} {arXiv:1801.04268 [astro-ph.CO]}
  \BibitemShut {NoStop}%
\bibitem [{\citenamefont {Caprini}\ \emph {et~al.}(2020)\citenamefont {Caprini}
  \emph {et~al.}}]{Caprini:2019egz}%
  \BibitemOpen
  \bibfield  {author} {\bibinfo {author} {\bibfnamefont {C.}~\bibnamefont
  {Caprini}} \emph {et~al.},\ }\bibfield  {title} {\bibinfo {title} {{Detecting
  gravitational waves from cosmological phase transitions with LISA: an
  update}},\ }\href {https://doi.org/10.1088/1475-7516/2020/03/024} {\bibfield
  {journal} {\bibinfo  {journal} {JCAP}\ }\textbf {\bibinfo {volume} {2003}},\
  \bibinfo {pages} {024}},\ \Eprint {https://arxiv.org/abs/1910.13125}
  {arXiv:1910.13125 [astro-ph.CO]} \BibitemShut {NoStop}%
%%CITATION = ARXIV:1910.13125;%%
\bibitem [{\citenamefont {Athron}\ \emph {et~al.}(2023)\citenamefont {Athron},
  \citenamefont {Bal\'azs}, \citenamefont {Fowlie}, \citenamefont {Morris},\
  and\ \citenamefont {Wu}}]{Athron:2023xlk}%
  \BibitemOpen
  \bibfield  {author} {\bibinfo {author} {\bibfnamefont {P.}~\bibnamefont
  {Athron}}, \bibinfo {author} {\bibfnamefont {C.}~\bibnamefont {Bal\'azs}},
  \bibinfo {author} {\bibfnamefont {A.}~\bibnamefont {Fowlie}}, \bibinfo
  {author} {\bibfnamefont {L.}~\bibnamefont {Morris}},\ and\ \bibinfo {author}
  {\bibfnamefont {L.}~\bibnamefont {Wu}},\ }\bibfield  {title} {\bibinfo
  {title} {{Cosmological phase transitions: from perturbative particle physics
  to gravitational waves}},\ }\href@noop {} {\  (\bibinfo {year} {2023})},\
  \Eprint {https://arxiv.org/abs/2305.02357} {arXiv:2305.02357 [hep-ph]}
  \BibitemShut {NoStop}%
\bibitem [{\citenamefont {Bodeker}\ and\ \citenamefont
  {Moore}(2017)}]{Bodeker:2017cim}%
  \BibitemOpen
  \bibfield  {author} {\bibinfo {author} {\bibfnamefont {D.}~\bibnamefont
  {Bodeker}}\ and\ \bibinfo {author} {\bibfnamefont {G.~D.}\ \bibnamefont
  {Moore}},\ }\bibfield  {title} {\bibinfo {title} {{Electroweak Bubble Wall
  Speed Limit}},\ }\href {https://doi.org/10.1088/1475-7516/2017/05/025}
  {\bibfield  {journal} {\bibinfo  {journal} {JCAP}\ }\textbf {\bibinfo
  {volume} {05}},\ \bibinfo {pages} {025}},\ \Eprint
  {https://arxiv.org/abs/1703.08215} {arXiv:1703.08215 [hep-ph]} \BibitemShut
  {NoStop}%
\bibitem [{\citenamefont {H\"oche}\ \emph {et~al.}(2021)\citenamefont
  {H\"oche}, \citenamefont {Kozaczuk}, \citenamefont {Long}, \citenamefont
  {Turner},\ and\ \citenamefont {Wang}}]{Hoche:2020ysm}%
  \BibitemOpen
  \bibfield  {author} {\bibinfo {author} {\bibfnamefont {S.}~\bibnamefont
  {H\"oche}}, \bibinfo {author} {\bibfnamefont {J.}~\bibnamefont {Kozaczuk}},
  \bibinfo {author} {\bibfnamefont {A.~J.}\ \bibnamefont {Long}}, \bibinfo
  {author} {\bibfnamefont {J.}~\bibnamefont {Turner}},\ and\ \bibinfo {author}
  {\bibfnamefont {Y.}~\bibnamefont {Wang}},\ }\bibfield  {title} {\bibinfo
  {title} {{Towards an all-orders calculation of the electroweak bubble wall
  velocity}},\ }\href {https://doi.org/10.1088/1475-7516/2021/03/009}
  {\bibfield  {journal} {\bibinfo  {journal} {JCAP}\ }\textbf {\bibinfo
  {volume} {03}},\ \bibinfo {pages} {009}},\ \Eprint
  {https://arxiv.org/abs/2007.10343} {arXiv:2007.10343 [hep-ph]} \BibitemShut
  {NoStop}%
\bibitem [{\citenamefont {Azatov}\ and\ \citenamefont
  {Vanvlasselaer}(2021)}]{Azatov:2020ufh}%
  \BibitemOpen
  \bibfield  {author} {\bibinfo {author} {\bibfnamefont {A.}~\bibnamefont
  {Azatov}}\ and\ \bibinfo {author} {\bibfnamefont {M.}~\bibnamefont
  {Vanvlasselaer}},\ }\bibfield  {title} {\bibinfo {title} {{Bubble wall
  velocity: heavy physics effects}},\ }\href
  {https://doi.org/10.1088/1475-7516/2021/01/058} {\bibfield  {journal}
  {\bibinfo  {journal} {JCAP}\ }\textbf {\bibinfo {volume} {01}},\ \bibinfo
  {pages} {058}},\ \Eprint {https://arxiv.org/abs/2010.02590} {arXiv:2010.02590
  [hep-ph]} \BibitemShut {NoStop}%
\bibitem [{\citenamefont {Gouttenoire}\ \emph {et~al.}(2021)\citenamefont
  {Gouttenoire}, \citenamefont {Jinno},\ and\ \citenamefont
  {Sala}}]{Gouttenoire:2021kjv}%
  \BibitemOpen
  \bibfield  {author} {\bibinfo {author} {\bibfnamefont {Y.}~\bibnamefont
  {Gouttenoire}}, \bibinfo {author} {\bibfnamefont {R.}~\bibnamefont {Jinno}},\
  and\ \bibinfo {author} {\bibfnamefont {F.}~\bibnamefont {Sala}},\ }\bibfield
  {title} {\bibinfo {title} {{Friction pressure on relativistic bubble
  walls}},\ }\href@noop {} {\  (\bibinfo {year} {2021})},\ \Eprint
  {https://arxiv.org/abs/2112.07686} {arXiv:2112.07686 [hep-ph]} \BibitemShut
  {NoStop}%
\bibitem [{\citenamefont {Baldes}\ \emph {et~al.}(2023)\citenamefont {Baldes},
  \citenamefont {Dichtl}, \citenamefont {Gouttenoire},\ and\ \citenamefont
  {Sala}}]{Baldes:2023fsp}%
  \BibitemOpen
  \bibfield  {author} {\bibinfo {author} {\bibfnamefont {I.}~\bibnamefont
  {Baldes}}, \bibinfo {author} {\bibfnamefont {M.}~\bibnamefont {Dichtl}},
  \bibinfo {author} {\bibfnamefont {Y.}~\bibnamefont {Gouttenoire}},\ and\
  \bibinfo {author} {\bibfnamefont {F.}~\bibnamefont {Sala}},\ }\bibfield
  {title} {\bibinfo {title} {{Bubbletrons}},\ }\href@noop {} {\  (\bibinfo
  {year} {2023})},\ \Eprint {https://arxiv.org/abs/2306.15555}
  {arXiv:2306.15555 [hep-ph]} \BibitemShut {NoStop}%
\bibitem [{\citenamefont {Ai}(2023)}]{Ai:2023suz}%
  \BibitemOpen
  \bibfield  {author} {\bibinfo {author} {\bibfnamefont {W.-Y.}\ \bibnamefont
  {Ai}},\ }\bibfield  {title} {\bibinfo {title} {{Logarithmically divergent
  friction on ultrarelativistic bubble walls}},\ }\href@noop {} {\  (\bibinfo
  {year} {2023})},\ \Eprint {https://arxiv.org/abs/2308.10679}
  {arXiv:2308.10679 [hep-ph]} \BibitemShut {NoStop}%
\bibitem [{\citenamefont {Watkins}\ and\ \citenamefont
  {Widrow}(1992)}]{Watkins:1991zt}%
  \BibitemOpen
  \bibfield  {author} {\bibinfo {author} {\bibfnamefont {R.}~\bibnamefont
  {Watkins}}\ and\ \bibinfo {author} {\bibfnamefont {L.~M.}\ \bibnamefont
  {Widrow}},\ }\bibfield  {title} {\bibinfo {title} {{Aspects of reheating in
  first order inflation}},\ }\href
  {https://doi.org/10.1016/0550-3213(92)90362-F} {\bibfield  {journal}
  {\bibinfo  {journal} {Nucl. Phys.}\ }\textbf {\bibinfo {volume} {B374}},\
  \bibinfo {pages} {446} (\bibinfo {year} {1992})}\BibitemShut {NoStop}%
%%CITATION = NUPHA,B374,446;%%
\bibitem [{\citenamefont {Konstandin}\ and\ \citenamefont
  {Servant}(2011)}]{Konstandin:2011ds}%
  \BibitemOpen
  \bibfield  {author} {\bibinfo {author} {\bibfnamefont {T.}~\bibnamefont
  {Konstandin}}\ and\ \bibinfo {author} {\bibfnamefont {G.}~\bibnamefont
  {Servant}},\ }\bibfield  {title} {\bibinfo {title} {{Natural Cold
  Baryogenesis from Strongly Interacting Electroweak Symmetry Breaking}},\
  }\href {https://doi.org/10.1088/1475-7516/2011/07/024} {\bibfield  {journal}
  {\bibinfo  {journal} {JCAP}\ }\textbf {\bibinfo {volume} {07}},\ \bibinfo
  {pages} {024}},\ \Eprint {https://arxiv.org/abs/1104.4793} {arXiv:1104.4793
  [hep-ph]} \BibitemShut {NoStop}%
\bibitem [{\citenamefont {Falkowski}\ and\ \citenamefont
  {No}(2013)}]{Falkowski:2012fb}%
  \BibitemOpen
  \bibfield  {author} {\bibinfo {author} {\bibfnamefont {A.}~\bibnamefont
  {Falkowski}}\ and\ \bibinfo {author} {\bibfnamefont {J.~M.}\ \bibnamefont
  {No}},\ }\bibfield  {title} {\bibinfo {title} {{Non-thermal Dark Matter
  Production from the Electroweak Phase Transition: Multi-TeV WIMPs and
  'Baby-Zillas'}},\ }\href {https://doi.org/10.1007/JHEP02(2013)034} {\bibfield
   {journal} {\bibinfo  {journal} {JHEP}\ }\textbf {\bibinfo {volume} {02}},\
  \bibinfo {pages} {034}},\ \Eprint {https://arxiv.org/abs/1211.5615}
  {arXiv:1211.5615 [hep-ph]} \BibitemShut {NoStop}%
%%CITATION = ARXIV:1211.5615;%%
\bibitem [{\citenamefont {Shakya}(2023)}]{Shakya:2023kjf}%
  \BibitemOpen
  \bibfield  {author} {\bibinfo {author} {\bibfnamefont {B.}~\bibnamefont
  {Shakya}},\ }\bibfield  {title} {\bibinfo {title} {{Aspects of Particle
  Production from Bubble Dynamics at a First Order Phase Transition}},\
  }\href@noop {} {\  (\bibinfo {year} {2023})},\ \Eprint
  {https://arxiv.org/abs/2308.16224} {arXiv:2308.16224 [hep-ph]} \BibitemShut
  {NoStop}%
\bibitem [{\citenamefont {Katz}\ and\ \citenamefont
  {Riotto}(2016)}]{Katz:2016adq}%
  \BibitemOpen
  \bibfield  {author} {\bibinfo {author} {\bibfnamefont {A.}~\bibnamefont
  {Katz}}\ and\ \bibinfo {author} {\bibfnamefont {A.}~\bibnamefont {Riotto}},\
  }\bibfield  {title} {\bibinfo {title} {{Baryogenesis and Gravitational Waves
  from Runaway Bubble Collisions}},\ }\href
  {https://doi.org/10.1088/1475-7516/2016/11/011} {\bibfield  {journal}
  {\bibinfo  {journal} {JCAP}\ }\textbf {\bibinfo {volume} {11}},\ \bibinfo
  {pages} {011}},\ \Eprint {https://arxiv.org/abs/1608.00583} {arXiv:1608.00583
  [hep-ph]} \BibitemShut {NoStop}%
\bibitem [{\citenamefont {Giudice}\ \emph {et~al.}(2024)\citenamefont
  {Giudice}, \citenamefont {Lee}, \citenamefont {Pomarol},\ and\ \citenamefont
  {Shakya}}]{Giudice:2024tcp}%
  \BibitemOpen
  \bibfield  {author} {\bibinfo {author} {\bibfnamefont {G.~F.}\ \bibnamefont
  {Giudice}}, \bibinfo {author} {\bibfnamefont {H.~M.}\ \bibnamefont {Lee}},
  \bibinfo {author} {\bibfnamefont {A.}~\bibnamefont {Pomarol}},\ and\ \bibinfo
  {author} {\bibfnamefont {B.}~\bibnamefont {Shakya}},\ }\bibfield  {title}
  {\bibinfo {title} {{Nonthermal heavy dark matter from a first-order phase
  transition}},\ }\href {https://doi.org/10.1007/JHEP12(2024)190} {\bibfield
  {journal} {\bibinfo  {journal} {JHEP}\ }\textbf {\bibinfo {volume} {12}},\
  \bibinfo {pages} {190}},\ \Eprint {https://arxiv.org/abs/2403.03252}
  {arXiv:2403.03252 [hep-ph]} \BibitemShut {NoStop}%
\bibitem [{\citenamefont {Freese}\ and\ \citenamefont
  {Winkler}(2023)}]{Freese:2023fcr}%
  \BibitemOpen
  \bibfield  {author} {\bibinfo {author} {\bibfnamefont {K.}~\bibnamefont
  {Freese}}\ and\ \bibinfo {author} {\bibfnamefont {M.~W.}\ \bibnamefont
  {Winkler}},\ }\bibfield  {title} {\bibinfo {title} {{Dark matter and
  gravitational waves from a dark big bang}},\ }\href
  {https://doi.org/10.1103/PhysRevD.107.083522} {\bibfield  {journal} {\bibinfo
   {journal} {Phys. Rev. D}\ }\textbf {\bibinfo {volume} {107}},\ \bibinfo
  {pages} {083522} (\bibinfo {year} {2023})},\ \Eprint
  {https://arxiv.org/abs/2302.11579} {arXiv:2302.11579 [astro-ph.CO]}
  \BibitemShut {NoStop}%
\bibitem [{\citenamefont {Cataldi}\ and\ \citenamefont
  {Shakya}(2024)}]{Cataldi:2024pgt}%
  \BibitemOpen
  \bibfield  {author} {\bibinfo {author} {\bibfnamefont {M.}~\bibnamefont
  {Cataldi}}\ and\ \bibinfo {author} {\bibfnamefont {B.}~\bibnamefont
  {Shakya}},\ }\bibfield  {title} {\bibinfo {title} {{Leptogenesis via bubble
  collisions}},\ }\href {https://doi.org/10.1088/1475-7516/2024/11/047}
  {\bibfield  {journal} {\bibinfo  {journal} {JCAP}\ }\textbf {\bibinfo
  {volume} {11}},\ \bibinfo {pages} {047}},\ \Eprint
  {https://arxiv.org/abs/2407.16747} {arXiv:2407.16747 [hep-ph]} \BibitemShut
  {NoStop}%
\bibitem [{\citenamefont {Cutting}\ \emph {et~al.}(2020)\citenamefont
  {Cutting}, \citenamefont {Escartin}, \citenamefont {Hindmarsh},\ and\
  \citenamefont {Weir}}]{Cutting:2020nla}%
  \BibitemOpen
  \bibfield  {author} {\bibinfo {author} {\bibfnamefont {D.}~\bibnamefont
  {Cutting}}, \bibinfo {author} {\bibfnamefont {E.~G.}\ \bibnamefont
  {Escartin}}, \bibinfo {author} {\bibfnamefont {M.}~\bibnamefont
  {Hindmarsh}},\ and\ \bibinfo {author} {\bibfnamefont {D.~J.}\ \bibnamefont
  {Weir}},\ }\bibfield  {title} {\bibinfo {title} {{Gravitational waves from
  vacuum first order phase transitions II: from thin to thick walls}},\
  }\href@noop {} {\  (\bibinfo {year} {2020})},\ \Eprint
  {https://arxiv.org/abs/2005.13537} {arXiv:2005.13537 [astro-ph.CO]}
  \BibitemShut {NoStop}%
\bibitem [{\citenamefont {Jinno}\ \emph {et~al.}(2019)\citenamefont {Jinno},
  \citenamefont {Konstandin},\ and\ \citenamefont {Takimoto}}]{Jinno_2019}%
  \BibitemOpen
  \bibfield  {author} {\bibinfo {author} {\bibfnamefont {R.}~\bibnamefont
  {Jinno}}, \bibinfo {author} {\bibfnamefont {T.}~\bibnamefont {Konstandin}},\
  and\ \bibinfo {author} {\bibfnamefont {M.}~\bibnamefont {Takimoto}},\
  }\bibfield  {title} {\bibinfo {title} {Relativistic bubble
  collisions{\textemdash}a closer look},\ }\href
  {https://doi.org/10.1088/1475-7516/2019/09/035} {\bibfield  {journal}
  {\bibinfo  {journal} {Journal of Cosmology and Astroparticle Physics}\
  }\textbf {\bibinfo {volume} {2019}}\bibinfo  {number} { (09)},\ \bibinfo
  {pages} {035}}\BibitemShut {NoStop}%
\bibitem [{\citenamefont {Inomata}\ \emph {et~al.}(2024)\citenamefont
  {Inomata}, \citenamefont {Kamionkowski}, \citenamefont {Kasai},\ and\
  \citenamefont {Shakya}}]{Inomata:2024rkt}%
  \BibitemOpen
\bibfield  {number} {  }\bibfield  {author} {\bibinfo {author} {\bibfnamefont
  {K.}~\bibnamefont {Inomata}}, \bibinfo {author} {\bibfnamefont
  {M.}~\bibnamefont {Kamionkowski}}, \bibinfo {author} {\bibfnamefont
  {K.}~\bibnamefont {Kasai}},\ and\ \bibinfo {author} {\bibfnamefont
  {B.}~\bibnamefont {Shakya}},\ }\bibfield  {title} {\bibinfo {title}
  {{Gravitational Waves from Particles Produced from Bubble Collisions in
  First-Order Phase Transitions}},\ }\href@noop {} {\  (\bibinfo {year}
  {2024})},\ \Eprint {https://arxiv.org/abs/2412.17912} {arXiv:2412.17912
  [astro-ph.CO]} \BibitemShut {NoStop}%
\end{thebibliography}%

\end{document}